 \documentclass[aps,draft,eqsecnum]{revtex4}
 \usepackage{epsf}  

\begin{document}


\title{Adjusted ADM systems and their expected stability properties:\\
constraint propagation analysis in Schwarzschild spacetime}
\author{Hisa-aki Shinkai}\email{shinkai@atlas.riken.go.jp}
\author{Gen Yoneda}\email{yoneda@mn.waseda.ac.jp}
\affiliation{
${}^\ast$ Computational Science Division,
Institute of Physical \& Chemical Research (RIKEN), \\
Hirosawa 2-1, Wako, Saitama, 351-0198 Japan
\\
${}^\dagger$ Department of Mathematical Sciences, Waseda University,
Shinjuku, Tokyo,  169-8555, Japan}

\date{January 7, 2002 (revised version), gr-qc/0110008}
\begin{abstract}

In order to find a way to have a better formulation for numerical
evolution of the Einstein equations, 
we study the propagation equations of the
constraints based on the Arnowitt-Deser-Misner formulation.
By adjusting constraint terms in the evolution equations, 
we try to construct an ``asymptotically constrained system"
which is expected to be robust against violation of the constraints,
and to enable 
a long-term stable and accurate numerical simulation. 
We first provide useful expressions for analyzing constraint propagation in a 
general spacetime, then apply it to Schwarzschild spacetime. 
We search when and where the negative real or non-zero imaginary 
eigenvalues of the homogenized
constraint propagation matrix appear, and how they depend on the choice
of coordinate system and adjustments.  
Our analysis includes the proposal of
Detweiler (1987), which is still the best one according to our conjecture
 but has a growing mode of error near the horizon.   
Some examples are snapshots
of a maximally sliced Schwarzschild black hole. 
The predictions here may help the community to make further improvements. 
\end{abstract}
\pacs{04.20.-q, 04.20.Fy, 04.25.-g 
and 04.25.Dm}

\maketitle

\section{Introduction}

Computing the Einstein equations numerically is a necessary direction
for general relativity research.  This approach, so-called numerical 
relativity, already gives us much feedback to help understand the nature of 
strong 
gravity, such as singularity formations, critical 
behavior of gravitational 
collapses, cosmology, and so on.  Numerical relativity will definitely be 
required in unveiling gravitational wave phenomena, highly relativistic 
astrophysical objects, and might be the only tool to discuss different 
gravitational theories or models \cite{reviewLehner}. 

There are several different approaches to simulate the Einstein equations, 
among them the most robust is to apply 3+1 (space + time) decomposition 
of spacetime, which was first formulated by Arnowitt, Deser and 
Misner (ADM) \cite{ADM} (we refer to this as the ``original ADM" system). 
A current important issue in the 3+1 approach is to control long-term 
stability of time integrations in 
simulating black hole or neutron star binary coalescences. 
So far, the use of fundamental variables $(g_{ij}, K_{ij})$, 
the internal 3-metric and extrinsic curvature, 
 has been the most popular approach \cite{ADM-York}
(we refer to this as the ``standard ADM" system). 
In recent years, however, many groups report that another
re-formulation of the Einstein equations provides 
more stable and accurate simulations.  
We try here to understand these efforts through
 ``adjusting" procedures to the evolution equations (that we 
explain later)
using eigenvalue analysis of the constraint propagation equations. 

At this moment, 
there may be three major directions to obtain longer time evolutions. 
The first is to use a modification of the ADM system that was developed
by Shibata and Nakamura \cite{SN} (and later remodified by 
Baumgarte and Shapiro \cite{BS}). 
This is a combination of the introduction of new variables, conformal
decompositions, rescaling the conformal factor, 
and the replacement of terms in the evolution equation using
momentum  constraints.  
Although there are many studies to show why this re-formulation 
is better than the standard ADM, as far as we know, there is no 
definite conclusion yet.  One observation \cite{potsdam0003} pointed out 
that to replace terms using momentum constraints appears to change the stability 
properties. 

The second direction is to re-formulate the Einstein equations in a first-order 
hyperbolic form \cite{hypref} \footnote{We note that rich
mathematical theories on PDEs are
obtained in a first-order form, while there is a study on linearized ADM
equations in a second-order form \cite{KreissOrtiz}.}.   
This is motivated from the
expectation that the symmetric hyperbolic system has well-posed properties in its Cauchy
treatment in many systems and also that the boundary treatment can be  improved if we
know the characteristic speed of the system.  We note that, in constructing
hyperbolic systems, the essential  procedures are to adjust equations
using constraints and to introduce new variables, normally the spatially 
derivatived metric. 
Several groups report that the hyperbolic formulation actually 
has advantages over the direct use of ADM formulation
\cite{BM,Cordoba1D,PennState_Sch1D,Cornell_Sch3D}. 
However it is also reported that there are no
drastic changes in the evolution properties {\it between} 
hyperbolic systems (weakly/strongly and symmetric hyperbolicity)
by systematic numerical studies 
by Hern \cite{HernPHD} based on Frittelli-Reula formulation \cite{FR96}, and
by the authors \cite{ronbun1} based on Ashtekar's formulation 
\cite{Ashtekar,ysPRL,ysIJMPD}\footnote{
\cite{PennState_prep} reports that 
there is an apparent difference on numerical stability
between these hyperbolicity levels, 
based on Kidder-Scheel-Teukolsky's reformulation \cite{Cornell_Sch3D}
of Anderson-York \cite{AndersonYork} and Frittelli-Reula \cite{FR96}
formulations.}.  
Therefore we may say that 
the mathematical notion of ``hyperbolicity" is not always applicable
for predicting the stability of numerical integration of the Einstein
equations (See also \S 4.4.2 in \cite{reviewLehner}).  
It will be useful if we have an alternative procedure
to predict stability 
including the effects of the non-principal parts of the equation, which are
neglected in the discussion of hyperbolicity.


The third is to construct a robust system against 
the violation of the constraints,
such that the constraint surface is the attractor.  
The idea was first proposed as ``$\lambda$-system" 
by Brodbeck et al \cite{BFHR} 
in which they introduce artificial flow to the constraint surface 
using a new variable 
based on the symmetric hyperbolic system. This idea was tested and 
 confirmed to 
work as expected in some test simulations by 
the authors\cite{ronbun2} (based on the formulation 
developed by \cite{SY-asympAsh}).  
Although it is questionable whether the
recovered solution is true evolution or 
not \cite{SiebelHuebner}, we think 
that to enforce the decay of errors in its initial 
perturbative stage is the key 
to the next improvements.  
Actually, by studying 
the evolution equations of the constraints 
(hereafter we call them constraint
propagation) and by evaluating eigenvalues 
(amplification factors, AFs) of constraint
propagation in its  homogenized form, we found that 
a similar ``asymptotically
constrained system" can be obtained by simply adjusting 
constraints to the
evolution equations, even  for the ADM equations \cite{ronbun3}.  

The purpose of this article is to extend 
our previous study \cite{ronbun3} in more
general expressions and also to apply the systems 
to 
spacetime which has non trivial curvature, 
Schwarzschild black hole spacetime. 
The actual numerical simulations require many ingredients to be
considered such as the choice of integration schemes, boundary treatments, 
grid structures and so on. 
However, we think that our approach to the stability problem through an 
{\it implementation to the equations} is definitely one of the aspects 
that should be improved. 

Adjusting evolution equations is not a new idea.  
Actually the standard ADM
system for numerical relativists \cite{ADM-York} is 
adjusted from the original one
\cite{ADM} using the Hamiltonian constraint (See
Frittelli's analysis on constraint propagation between the original and
standard ADM formulations \cite{Fri-con}).  
Detweiler \cite{detweiler} proposed a
system using adjustments so that the L2 norm of constraints may not 
blow up (we will study this later in detail). 
Several numerical relativity groups recently report 
the advantages of the adjusting procedure 
with a successful example 
\cite{PennState_Sch1D,Cordoba1D}. 
We  try here to understand the background mathematical features systematically 
by using AFs of constraint propagation.

We conjecture that we can construct
an asymptotically constrained system
by evaluating the AFs in advance.  
Especially,
we propose that if the amplification factors are negative or pure-imaginary,
then the system becomes more stable.  
This conjecture is already approved 
by numerical studies \cite{ronbun2} in the Maxwell equations 
and in the Ashtekar
version of the Einstein equations on the Minkowskii background.  
In this article, we show when and where such a feature
is available using the ADM equations for Schwarzschild spacetime.  
Although the analysis here is restricted to the fixed background, 
we believe that this study clarifies the
properties of adjustments and indicates further possibilities. 

In section \ref{Sec2}, we describe our idea of ``adjusted systems" 
generally, and
their application to the ADM formulation.  
We then explain how we calculate the ``amplification factors" in the spherically 
symmetric spacetime in \S \ref{Sec3},
and show several examples in \S \ref{Sec4}.
The concluding remarks are in \S \ref{Sec5}.
Appendix \ref{appA} summarizes the constraint propagation equations
of the ADM formulation generally, which may be useful for further studies. 
Appendix \ref{appB} explains briefly a way to obtain maximally 
sliced Schwarzschild spacetime that is used to obtain Fig.\ref{fig7max}. 

\section{Adjusted systems}\label{Sec2}
\subsection{Procedure and background -- general discussion --}\label{Sec2A}

We begin with an overview of the adjusting procedure and the idea of background 
structure, which were described in 
our previous work \cite{ronbun2,ronbun3}.  

Suppose we have a set of dynamical variables   $u^a (x^i,t)$, and their
evolution equations
\begin{equation}
\partial_t u^a = f(u^a, \partial_i u^a, \cdots), \label{ueq}
\end{equation}
and the
(first class) constraints
\begin{equation}
C^\alpha (u^a, \partial_i u^a, \cdots) \approx 0.
\end{equation}
For monitoring the violation of constraints, we propose to investigate
the evolution equation of $C^\alpha$ (constraint propagation),
\begin{equation}
\partial_t C^\alpha = g(C^\alpha, \partial_i C^\alpha, \cdots). \label{Ceq}
\end{equation}
(We do not mean to integrate (\ref{Ceq}) numerically, but to evaluate them
analytically in advance.)
Then, there may be two major analyses of (\ref{Ceq}); 
(a) the hyperbolicity of (\ref{Ceq})
if  (\ref{Ceq}) forms a first order form, and 
(b) the eigenvalue analysis of the whole RHS in 
(\ref{Ceq}) after certain adequate homogenized procedures. 

If the evolution equations form a first-order system, 
the standard hyperbolic PDE analysis is applicable. 
The analysis is mainly to 
identify the level of hyperbolicity
and to calculate the characteristic speed of the system,  from eigenvalues 
of the principal matrix.
We expect mathematically rigorous well-posed features for 
strongly hyperbolic or symmetric
hyperbolic systems.  Also we know the characteristic speeds suggest
the satisfactory criteria
for stable evolutions; such as when they are real,  under the propagation
speed of the  original variables, $u^a$,
and/or within the causal region of the numerical integration
scheme that is applied.

For example, 
the evolution equations of the standard/original ADM formulation, (\ref{ueq}) 
[(\ref{admevo1}) and (\ref{admevo2})], do not form a first-order system, while
their constraint propagation equations, (\ref{Ceq}) 
[(\ref{CHproADM}) and (\ref{CMproADM})], form a first-order system. 
Therefore, we can apply the classification on the hyperbolicity (weakly,
strongly or symmetric) to the constraint propagation equations. 
However, 
if one adjusts ADM equations with constraints, then this first-order 
characters will not be guaranteed. 
Another problem in the hyperbolic analysis is that it only 
discusses the principal part of the system, 
and ignores the rest.  If there is a method to characterize 
the non-principal part, then that will help to clarify 
our understanding of stability.  (This importance 
was mentioned in \cite{reviewLehner}.)

The second analysis, the eigenvalue analysis of the whole RHS in 
(\ref{Ceq}), may compensate for the above problems. 
We propose  to homogenize (\ref{Ceq}) by a Fourier transformation, 
e.g. 
\begin{eqnarray}
\partial_t \hat{C}^\alpha &=& \hat{g}(\hat{C}^\alpha)
=M^\alpha{}_{\beta} \hat{C}^\beta,
\nonumber \\ &&
\mbox{~~~where~}
C(x,t)^\rho
=\displaystyle{\int} \hat{C}(k,t)^\rho\exp(ik\cdot x)d^3k,
\label{CeqF}
\end{eqnarray}
then to analyze the set of eigenvalues, say $\Lambda$s,  of the 
coefficient matrix, $M^\alpha{}_{\beta}$, in
(\ref{CeqF}).  We call $\Lambda$s the amplification factors (AFs)
of (\ref{Ceq}). 
As we  have proposed and confirmed in \cite{ronbun2}:

\begin{quotation}
\noindent
{\bf Conjecture: } 
\begin{itemize}
\item[(a)] If the amplification factors have a {\it negative real-part } (the constraints 
are forced to be
diminished), then we see
more stable evolutions than a system which has positive amplification factors.
\item[(b)] If the amplification factors have a {\it non-zero  imaginary-part }
(the constraints are propagating away), then we see
more stable evolutions than a system which has 
zero amplification factors.
\end{itemize}
\end{quotation}
We found heuristically that the system becomes more stable when
more $\Lambda$s satisfy the above criteria \cite{ronbun1,ronbun2} 
\footnote{
We note that the 
{\it non-zero} eigenvalue feature was conjectured 
in Alcubierre {\it et. al.} \cite{potsdam0003} in order to show the 
advantage of the conformally-scaled ADM
system,  but the discussion there is of linearized dynamical equations and not of 
constraint propagation equations. }.  
We remark that this eigenvalue analysis requires the fixing of
a particular background spacetime,  
since the AFs depend on the dynamical variables, $u^a$.

The above features of the constraint propagation, (\ref{Ceq}),
will differ when we modify the original evolution equations.
Suppose we add (adjust) the evolution equations using constraints
\begin{equation}
\partial_t u^a = f(u^a, \partial_i u^a, \cdots)
+ F(C^\alpha, \partial_i C^\alpha, \cdots), \label{DeqADJ}
\end{equation}
then (\ref{Ceq}) will also be modified as
\begin{equation}
\partial_t C^\alpha = g(C^\alpha, \partial_i C^\alpha, \cdots)
+ G(C^\alpha, \partial_i C^\alpha, \cdots). \label{CeqADJ}
\end{equation}
Therefore, the problem is how to adjust the evolution equations
so that their constraint propagations satisfy the above criteria 
as much as possible. 

\subsection{Standard ADM system and its constraint propagation}
We start by analyzing the standard ADM system.
By ``standard ADM" we mean here the most widely adopted system, due to
York \cite{ADM-York},
with the evolution equations,
\begin{eqnarray}
\partial_t \gamma_{ij}&=&
-2\alpha K_{ij}+\nabla_i\beta_j+\nabla_j\beta_i, \label{admevo1}
\\
\partial_t K_{ij} &=&
\alpha R^{(3)}_{ij}+\alpha K K_{ij}-2\alpha K_{ik}{K^k}_j
-\nabla_i\nabla_j \alpha
\nonumber \\ &&
+(\nabla_i \beta^k) K_{kj} +(\nabla_j \beta^k) K_{ki}
+\beta^k \nabla_k K_{ij},  \label{admevo2}
\end{eqnarray}
and the constraint equations,
\begin{eqnarray}
{\cal H}&:=& R^{(3)}+K^2-K_{ij}K^{ij}, \label{admCH} 
\\
{\cal M}_i&:=& \nabla_j {K^j}_i-\nabla_i K,  \label{admCM}
\end{eqnarray}
where $(\gamma_{ij}, K_{ij})$ are the induced three-metric
and the extrinsic
curvature, $(\alpha, \beta_i)$ are the lapse function
and the shift covector,
$\nabla_i$ is the covariant derivative adapted to $\gamma_{ij}$,
and $R^{(3)}_{ij}$ is the three-Ricci tensor.

The constraint propagation equations,
which are the time evolution equations
of the Hamiltonian constraint (\ref{admCH}) and
the momentum constraints (\ref{admCM}), can be written as
\begin{eqnarray}
\partial_t {\cal H}&=&
\beta^j (\partial_j {\cal H})
+2\alpha K{\cal H}
-2\alpha \gamma^{ij}(\partial_i {\cal M}_j)
\nonumber \\ &&
+\alpha(\partial_l \gamma_{mk})
   (2\gamma^{ml}\gamma^{kj}-\gamma^{mk}\gamma^{lj}){\cal M}_j
-4\gamma^{ij} (\partial_j\alpha){\cal M}_i,
\label{CHproADM}
\\
\partial_t {\cal M}_i&=&
-(1/2)\alpha (\partial_i {\cal H})
-(\partial_i\alpha){\cal H}
+\beta^j (\partial_j {\cal M}_i)
\nonumber \\ &&
+\alpha K {\cal M}_i
-\beta^k\gamma^{jl}(\partial_i\gamma_{lk}){\cal M}_j
+(\partial_i\beta_k)\gamma^{kj}{\cal M}_j.
\label{CMproADM}
\end{eqnarray}
Further expressions of these constraint propagations are introduced
in Appendix \ref{appA} of this article. 

\subsection{Adjustment to ADM evolution equations and its effects on
constraint propagations} 
Generally, we can write the adjustment terms to
(\ref{admevo1}) and (\ref{admevo2})
using (\ref{admCH}) and (\ref{admCM}) by the following combinations
(using up to the first derivatives of constraints for simplicity
in order to include Detweiler's case, see the next subsection),
\begin{eqnarray}
&\mbox{adjustment term of }\quad
\partial_t \gamma_{ij}:& \quad
+P_{ij} {\cal H}
+Q^k{}_{ij}{\cal M}_k
+p^k{}_{ij}(\nabla_k {\cal H})
+q^{kl}{}_{ij}(\nabla_k {\cal M}_l), \label{adjADM1}
\\
&\mbox{adjustment term of}\quad
\partial_t K_{ij}:& \quad
+R_{ij} {\cal H}
+S^k{}_{ij}{\cal M}_k
+r^k{}_{ij} (\nabla_k{\cal H})
+s^{kl}{}_{ij}(\nabla_k {\cal M}_l), \label{adjADM2}
\end{eqnarray}
where $P, Q, R, S$ and $p, q, r, s$ are multipliers  (please do not
confuse $R_{ij}$ with
three Ricci curvature  that we write as $R^{(3)}_{ij}$).

According to this adjustment, the constraint propagation equations are 
also modified as
\begin{eqnarray}
\partial_t {\cal H} &=&
\mbox{(\ref{CHproADM})}
  + H_1^{mn}  (\ref{adjADM1})
+H_2^{imn}\partial_i(\ref{adjADM1})
+H_3^{ijmn}\partial_i\partial_j (\ref{adjADM1})
+H_4^{mn}(\ref{adjADM2}), \label{Hnewhosei}
\\
\partial_t {\cal M}_i &=&
\mbox{(\ref{CMproADM})}
+ M_1{}_i{}^{mn} (\ref{adjADM1})
+M_2{}_i{}^{jmn} \partial_j(\ref{adjADM1})
+M_3{}_i{}^{mn} (\ref{adjADM2})
+M_4{}_i{}^{jmn} \partial_j(\ref{adjADM2}). \label{Mnewhosei}
\end{eqnarray}
with appropriate changes in indices.  (See detail in 
Appendix \ref{appA}. The definitions of
 $H_1, \cdots, M_1, \cdots$ are also there.)

\subsection{Examples of adjustments}
We show several examples of adjustments here. 

The first test is to show the differences between 
the standard ADM \cite{ADM-York}
and the original ADM system \cite{ADM}.  
In
terms of (\ref{adjADM1}) and (\ref{adjADM2}), the adjustment, 
\begin{eqnarray}
R_{ij}=\kappa_F \alpha \gamma_{ij}, 
\label{originalADMadjust}
\end{eqnarray}
will distinguish two, 
where $\kappa_F$ is a constant
and set the other multipliers zero. 
Here $\kappa_F=0$ corresponds to the
standard ADM (no adjustment),
and $\kappa_F=-1/4$ to the original ADM (without any
adjustment to the canonical formulation by ADM).
As one can check the adjusted constraint propagation equations 
in Appendix \ref{appA}, 
adding $R_{ij}$ term keeps the constraint propagation in a 
first-order form.  Frittelli \cite{Fri-con} (see also \cite{ronbun3}) 
pointed out that
the  hyperbolicity of constraint propagation equations is better 
in the standard ADM system. 

The second example is the one proposed by Detweiler \cite{detweiler}. 
He found that with a particular combination (see below), 
 the evolution of the
energy norm of the constraints, ${\cal H}^2+{\cal M}^2$,
can be negative definite
when we apply the maximal slicing condition, $K=0$.
His adjustment can be written in our notation in
(\ref{adjADM1}) and (\ref{adjADM2}), as
\begin{eqnarray}
P_{ij}&=&-  \kappa_L \alpha^3 \gamma_{ij}, \label{Det1}\\
R_{ij}&=& \kappa_L \alpha^3 (K_{ij}- (1 / 3) K \gamma_{ij}), \label{Det2}
\\
S^k{}_{ij}&=& \kappa_L \alpha^2 [
    {3}  (\partial_{(i} \alpha) \delta_{j)}^{k}
- (\partial_l \alpha) \gamma_{ij} \gamma^{kl}], \label{Det3}
\\
s^{kl}{}_{ij}&=&\kappa_L \alpha^3 [
  \delta^k_{(i}\delta^l_{j)}-(1/3) \gamma_{ij}\gamma^{kl}],
  \label{Det4}
\end{eqnarray}
everything else is zero, where $\kappa_L$ is a constant.
Detweiler's adjustment, (\ref{Det1})-(\ref{Det4}),
does not put constraint propagation equation
to a first order form, so we cannot discuss
hyperbolicity  or the characteristic speed of the constraints.
We confirmed numerically, using perturbation on Minkowskii spacetime,
that Detweiler's system provides better
accuracy than the standard ADM, but only for small positive $\kappa_L$.
See the Appendix of \cite{ronbun3}.

There are infinite ways of adjusting equations.
Actually one of our criteria, the negative real AFs, requires breaking
the time-symmetric features of the standard ADM system (when we 
apply them to the time-symmetric background metric). 
One observation by us \cite{ronbun3} is that such AFs are available 
if we adjust the terms which break the 
time reversal symmetry of the evolution equations. That is, 
by reversing the time ($\partial_t \rightarrow -\partial_t$), there are
variables which change their signatures
 [$K_{ij}, \partial_t \gamma_{ij}, {\cal M}_i, \cdots$], 
while not [$g_{ij}, \partial_t K_{ij}, {\cal H}, \cdots$]. 
(Here we are
assuming the 3-metric $\gamma_{ij}$ has plus 
parity for time reversal symmetry). 
So that the multiplier, for example,  $P_{ij}$
in (\ref{adjADM1})  is desired to have plus parity in order to 
break the minus parity of 
$\partial_t \gamma_{ij}$ equation.  

We list several combinations of adjustments in Table \ref{table1}
and Table \ref{table3}. (Table \ref{table1} shows the ones we have plotted,
and the list of plots is in Table \ref{table2}).
We defined the adjustment terms so that their positive 
multiplier parameter, $\kappa > 0$, makes the system 
{\it better} in stability according to our conjecture.  
(Here {\it better} means in accordance with 
our conjecture in \S \ref{Sec2A}). 
In this article, we do not discuss the ranges of the effective
multiplier parameter, $\kappa$, since the range depends on the
characteristic speeds of the models and 
numerical integration schemes as we observed in \cite{ronbun2}.

We show how these adjustments change the AFs in Schwarzschild spacetime
in \S \ref{Sec4}. 

\section{Constraint propagations in spherically symmetric spacetime}
\label{Sec3}
{}From here, we restrict our discussion
 to spherically symmetric spacetime.  
We introduce the violation of constraints as its perturbation 
using harmonics. 
According to our motivation, the actual procedure to analyze the 
adjustments is to substitute the
perturbed metric to the (adjusted) evolution equations first and to 
evaluate the according
perturbative errors in the (adjusted) constraint propagation equations. 
However, for the simplicity, we apply the perturbation to the pair of
constraints directly and analyze the effects of adjustments in its 
propagation equations.  
The latter, we think, presents the feature of constraint propagation more 
clearly for our purposes.

\subsection{The procedure}
The discussion becomes clear if we expand the constraint 
$C_\mu := ({\cal H}, {\cal M}_i)^T$
using vector harmonics,  
\begin{equation}
C_\mu = \sum_{l,m} \left(  A^{lm} a_{lm} + B^{lm} b_{lm} + C^{lm} c_{lm} + 
D^{lm} d_{lm} \right), 
\label{vectorharmonics}
\end{equation}
where we choose the basis as 
\begin{eqnarray}
{a}_{lm} (\theta, \varphi) &=&(Y_{lm},0,0,0)^T, \label{vectorharmonics1}
\\
{b}_{lm} (\theta, \varphi) &=&(0,Y_{lm},0,0)^T,
\\
{c}_{lm} (\theta, \varphi) &=&
{r \over \sqrt{ l (l+1)}}( 0,  0, \partial_\theta Y_{lm} , \partial_\varphi 
Y_{lm})^T, 
\\
{d}_{lm} (\theta, \varphi) &=&
{r \over \sqrt{ l (l+1)}}
(0, 0,  - {1 \over \sin \theta} \partial_\varphi Y_{lm} , \sin \theta
\, \partial_\theta Y_{lm} )^T,  \label{vectorharmonics4}
\end{eqnarray}
and the  coefficients $A^{lm}, \cdots, D^{lm}$ are functions of $(t, r)$.
Here $Y_{lm}$ is the spherical harmonic function, 
\begin{equation}
Y_{lm}(\theta,\varphi) =  
 (-1)^{(m+|m|)/2} \sqrt{   {(2l+1)\over 4 
\pi}{(l-|m|)! \over  (l+|m|)!}} \, 
P^m_l (\cos \theta) e^{im \varphi}. 
\end{equation}
The basis (\ref{vectorharmonics1})-(\ref{vectorharmonics4}) are normalized 
so that they satisfy
\begin{equation}
\langle C_\mu, C_\nu \rangle = \int_0^{2 \pi} d \varphi \int_0^\pi 
\,  C^\ast_\mu C_\rho \,  \eta^{\mu \rho} 
\sin \theta d \theta,
\end{equation}
where $\eta^{\mu \rho}$ is Minkowskii metric and the asterisk denotes 
the complex conjugate. 
Therefore
\begin{equation}
A^{lm} =\langle a^{lm}_{(\nu)}, C_\nu \rangle,  \quad 
\partial_t A^{lm} =\langle a^{lm}_{(\nu)}, \partial_t C_\nu \rangle, 
\quad \mbox{etc.}
\end{equation}

In order to analyze the radial dependences, 
we also express these evolution equations using the Fourier expansion
on the radial coordinate, 
\begin{equation}
A^{lm} = \sum_k \hat A^{lm}_{(k)}(t) \, e^{ik r} \quad \mbox{etc.}
\end{equation}
So that we can obtain the RHS of the evolution equations
for $(\hat A^{lm}_{(k)}(t), \cdots, \hat D^{lm}_{(k)}(t))^T$
in a homogeneous form.

\subsection{Expression for the standard ADM formulation}

We write the spherically symmetric spacetime using a metric, 
\begin{eqnarray}
ds^2 &=& -(\alpha(t,r)^2-\beta_r(t,r)^2/\gamma_{rr})dt^2
+2\beta_r(t,r)dtdr
\nonumber \\ &&
+\gamma_{rr}(t,r)dr^2
+\gamma_{\theta\theta}(t,r)(d\theta^2+\sin^2\theta d\varphi^2), 
\end{eqnarray}
where $\alpha$ and $\beta_r$ are the lapse function and the shift vector. 
$\gamma_{rr}$ and $\gamma_{\theta\theta}$ are also interpreted as 3-metric in 3+1
decomposition, so that the relevant extrinsic curvature, $K_{ij}$, become
$
K_{ij}=\mbox{diag}(K_{rr},K_{\theta\theta},K_{\theta\theta}\sin^2\theta)$ and
its trace becomes $K=K_{rr}/\gamma_{rr}+2K_{\theta\theta}/\gamma_{\theta\theta}. $

According to the procedure in the previous section, we obtain the 
constraint propagation equations for the standard ADM formulation, 
(\ref{CHproADM}) and (\ref{CMproADM}), in the following form:
\begin{eqnarray}
\partial_t A^{lm} (t,r) &=&
2\alpha K A^{lm}
+\beta^r (\partial_r A^{lm})
\nonumber \\&&
-2\alpha\gamma^{rr}
(\partial_r B^{lm})
+\big(\alpha(\partial_r \gamma_{rr})\gamma^{rr}\gamma^{rr}
-2\alpha(\partial_r \gamma_{\theta\theta})\gamma^{\theta\theta}\gamma^{rr}
-4\gamma^{rr} (\partial_r\alpha)\big)B^{lm}
\nonumber \\&&
-2\alpha\gamma^{\theta \theta}
r \sqrt{ l (l+1)}
C^{lm},  \label{standardADMconpro1}
\\
\partial_t B^{lm} (t,r) &=&
-(1/2)\alpha\partial_r A^{lm}
-(\partial_r\alpha)A^{lm}
\nonumber \\&&
+(\alpha K
-\beta^r\gamma^{rr}(\partial_r\gamma_{rr})
+(\partial_r\beta_r)\gamma^{rr})B^{lm}
+\beta^r \partial_r B^{lm},   \label{standardADMconpro2}
\\
\partial_t C^{lm} (t,r) &=&
-{\alpha\sqrt{l(l+1)} \over 2r}
A^{lm}
+\alpha K C^{lm}
+\beta^r(
{1\over r} C^{lm}
+  \partial_r C^{lm}  ),  \label{standardADMconpro3}
\\
\partial_t D^{lm} (t,r) &=&
\alpha K D^{lm}
+\beta^r
({1\over r} D^{lm}+
\partial_r D^{lm}),  \label{standardADMconpro4}
\end{eqnarray}
and then 
\begin{eqnarray}
{d\over dt} \hat{A}^{lm}_{(k)} (t) &=&
(2\alpha K +ik \beta^r ) \hat{A}^{lm}_{(k)}
\nonumber \\&&
+\big(
-2ik \alpha\gamma^{rr}
+\alpha(\partial_r \gamma_{rr})\gamma^{rr}\gamma^{rr}
-2\alpha(\partial_r \gamma_{\theta\theta})\gamma^{\theta\theta}\gamma^{rr}
-4\gamma^{rr} (\partial_r\alpha)\big) \hat{B}^{lm}_{(k)}
\nonumber \\&&
-2\alpha\gamma^{\theta \theta}r \sqrt{ l (l+1)}  \hat{C}^{lm}_{(k)}, 
\label{standardADMconproF1}
\\
{d\over dt} \hat{B}^{lm}_{(k)} (t) &=&
(-(ik/2)\alpha -(\partial_r\alpha))\hat{A}^{lm}_{(k)}
+(\alpha K
-\beta^r\gamma^{rr}(\partial_r\gamma_{rr})
+(\partial_r\beta_r)\gamma^{rr}
+ik\beta^r ) \hat{B}^{lm}_{(k)}, 
\label{standardADMconproF2}
\\
{d\over dt} \hat{C}^{lm}_{(k)} (t)&=&
-{\alpha\sqrt{l(l+1)} \over 2r}\hat{A}^{lm}_{(k)}
+(\alpha K+{\beta^r\over r}   +  ik\beta^r )\hat{C}^{lm}_{(k)}, 
\label{standardADMconproF3}
\\
{d\over dt} \hat{D}^{lm}_{(k)} (t) &=&
(\alpha K
+{\beta^r\over r} +
ik \beta^r)\hat{D}^{lm}_{(k)}. 
\label{standardADMconproF4}
\end{eqnarray}
There is no dependence on $m$. 
We see that the expressions are equivalent to the case of
flat background spacetime \cite{ronbun3} when 
we take $l=0$ and $r \rightarrow \infty$. 
Therefore our results also show the behavior of the flat background limit
in its large $r$ limit. 

\subsection{Example: original ADM formulation}
We only present here one example, the comparison between the 
standard and original ADM systems. 
By substituting (\ref{originalADMadjust}) into 
(\ref{Hnewhosei}) and (\ref{Mnewhosei}), we obtain 
\begin{eqnarray}
\partial_t {\cal H}
&=& (\ref{CHproADM}) + 4 \kappa_1 \alpha K  {\cal H}, 
\\
\partial_t {\cal M}_i
&=& (\ref{CMproADM}) - 2  \kappa_1 \alpha (\partial_i {\cal H})
-2\kappa_1 (\partial_i\alpha) {\cal H}. 
\end{eqnarray}
In the spherically symmetric spacetime, 
this adjustment affects 
(\ref{standardADMconpro1})-(\ref{standardADMconpro4}) as follows:
\begin{eqnarray}
\partial_t A^{lm}(t,r)&=&
(\ref{standardADMconpro1}) + 4 \kappa_1 \alpha K A^{lm}, 
\label{originalADMconpro1}
\\
\partial_t B^{lm}(t,r)&=&
(\ref{standardADMconpro2}) - 2 \kappa_1 \alpha \partial_r A^{lm}
-2\kappa_1 (\partial_r\alpha)A^{lm}, 
\label{originalADMconpro2}
\\
\partial_t C^{lm}(t,r)&=&
(\ref{standardADMconpro3})
-{2\kappa_1 \alpha\sqrt{l(l+1)} \over r} A^{lm}, 
\label{originalADMconpro3}
\\
\partial_t D^{lm}(t,r)&=&
(\ref{standardADMconpro4}). 
\label{originalADMconpro4}
\end{eqnarray}
After homogenization, 
(\ref{standardADMconproF1})-(\ref{standardADMconproF4}) become
\begin{eqnarray}
{d \over dt}
\left(\matrix{\hat{A}^{lm}_{(k)} \cr \hat{B}^{lm}_{(k)} \cr 
\hat{C}^{lm}_{(k)} \cr \hat{D}^{lm}_{(k)}}\right)
=
\left(\matrix{(\ref{standardADMconproF1}) \cr (\ref{standardADMconproF2}) \cr 
(\ref{standardADMconproF3}) \cr (\ref{standardADMconproF4})} \right)
+
\left(\matrix{
4\kappa_1 \alpha K &0&0&0 \cr
-2\kappa_1 \alpha ik-2\kappa_1 (\partial_r \alpha) &0&0&0 \cr
-{2\kappa_1 \alpha\sqrt{l(l+1)} / r}&0&0&0 \cr
0&0&0&0 \cr
}\right)
\left(\matrix{\hat{A}^{lm}_{(k)} \cr \hat{B}^{lm}_{(k)} 
\cr \hat{C}^{lm}_{(k)} \cr \hat{D}^{lm}_{(k)}}\right). \label{326}
\end{eqnarray}
The  eigenvalues (AFs) $\Lambda^i$ of
the RHS matrix of (\ref{326}) 
can be calculated by fixing the metric components including the gauge. 
For example, 
on the standard Schwarzschild metric (\ref{standardSch}), they are
\begin{eqnarray}
\Lambda^i &=&(0,0,\sqrt{a},-\sqrt{a}) \nonumber \\
a&=& -k^2 + { 4M k^2 r^2 (r-M)+ 2M(2r-M) + l(l+1) r (r-2M) + ikr(2r^2-3Mr-2M^2) 
\over r^4} \label{eigenvalueADM}
\end{eqnarray}
for the choice of $\kappa_1=0$ (the standard ADM), while they are
\begin{equation}
\Lambda^i =(0,0,\sqrt{b},-\sqrt{b}), \qquad
b= {M (2r-M) + irkM (2M-r) \over r^4} \label{eigenvalueOADM}
\end{equation}
for the choice of $\kappa_1=-1/4$ (the original ADM).

\subsection{Our analytic approach}
The above example is the simplest one.  
In the next section, we will show the AFs of the adjusted systems
shown in Table \ref{table1}. 
We found that to write down  
the analytical expressions of them is not a good idea due to their 
length. 
We will therefore  plot AFs to see if the real parts become negative, 
or if the imaginary
parts become non-zero or not.
 
We used the symbolic calculation software, 
{\it Mathematica} and {\it Maple}, and made plots by checking two 
independent outputs. 
These scripts are
available upon request. 

\section{Constraint propagations in Schwarzschild spacetime}\label{Sec4}

\subsection{Coordinates}
We present our analysis of the constraint propagation equations
in Schwarzschild black hole spacetime, 
\begin{equation}
ds^2=-(1-{2M \over r})dt^2+{dr^2 \over 1-{2M/ r}} +r^2 d\Omega^2,   
\qquad\mbox{(the~standard~expression)} 
\label{standardSch}
\end{equation}
where $M$ is the mass of a black hole. 
For numerical relativists, evolving a single black hole
is the essential test problem, though it is a trivial at first sight. 
The standard expression, (\ref{standardSch}), has a coordinate 
singularity at $r=2M$, so that we need to move another coordinate
for actual numerical time integrations.  

One alternative is the isotropic coordinate, 
\begin{equation}
ds^2=-({1-{M/2 r_{iso}} \over 1+{M/2r_{iso}}})^2dt^2+
(1+{M\over 2r_{iso}})^4 [dr_{iso}^2+r_{iso}^2 d\Omega^2], 
\qquad\mbox{(the~isotropic~expression)} 
\label{isoSch}
\end{equation}
which is given by the coordinate transformation, 
$r=(1+{M / 2 r_{iso}})^2 r_{iso}$. 
Here $r_{iso}=M/2$ indicates the minimum throat radius of the 
Einstein-Rosen bridge.  
Bernstein, Hobill and Smarr \cite{BHS89} showed a systematic comparison for 
numerical integration schemes by 
applying the coordinate transformation, 
$r_{iso}={M } e^{r_{new}} / 2$,  further to (\ref{isoSch}). 

The expression of the ingoing Eddington-Finkelstein (iEF) coordinate has become
popular in numerical relativity for treating black hole boundaries
as an excision, since iEF
penetrates the horizon without a coordinate singularity.
The expression is, 
\begin{equation}
ds^2=-(1-{2M\over r})dt_{iEF}^2+{4M \over r} dt_{iEF} dr 
+ (1+{2M\over r}) dr^2
+r^2 d\Omega^2 
\qquad\mbox{(the~iEF~expression)} 
\label{iEFSch}
\end{equation}
which is given by $t_{iEF}=t+2M \log (r - 2M)$ and 
the radial coordinate is common to
(\ref{standardSch}).
The geometrical interpretation of the iEF coordinate system is that
in addition to having a
timelike killing vector, the combination of timelike and
radial tangent vectors $\vec\partial_t - \vec\partial_r$ remains null.

Another expression we test is the
Painlev\'e-Gullstrand (PG) 
coordinates, 
\begin{equation}
ds^2 = - \left( 1-\frac {2\,M}{r}\right) dt_{PG}^2
+ 2 \sqrt{\frac{2\,M}{r}} dt_{PG}\, dr
+ dr^2 + r^2 d\Omega ^2 \;,
\qquad\mbox{(the~PG~expression)} 
\label{PGSch}
\end{equation}
which is given by 
$t_{PG}=t+ \sqrt{8Mr} - 2M \log \{ (\sqrt{r/2M}+1) / (\sqrt{r/2M}-1) \}$ 
and the radial coordinate is common to
(\ref{standardSch}).
The PG coordinate system can be viewed as that
anchored to a family of freely moving observers (time-like)
starting at infinity with vanishing velocity \cite{Poisson}.

The latter two (iEF/PG) are also 
different from the former (standard/isotropic) two in the point
that their extrinsic curvature on the initial slice ($t_{iEF}=t_{PG}=0$) 
is not zero.  
To conclude first, the effects of adjustments are similar both
between the former  and between the latter.

In Table. \ref{table2}, we list the combinations 
(adjustments and coordinates) we plotted.

\subsection{in the standard Schwarzschild coordinate}
We show first the case of the standard Schwarzschild 
coordinate, (\ref{standardSch}), 
since this example provides a basic overview of our analysis.
The cases of the isotropic 
coordinate, (\ref{isoSch}), will be shown to be quite similar.  

\begin{figure}[p]
\setlength{\unitlength}{1cm}
\begin{picture}(15,11)
\put(1.0,5.5){\epsfxsize=6.0cm \epsffile{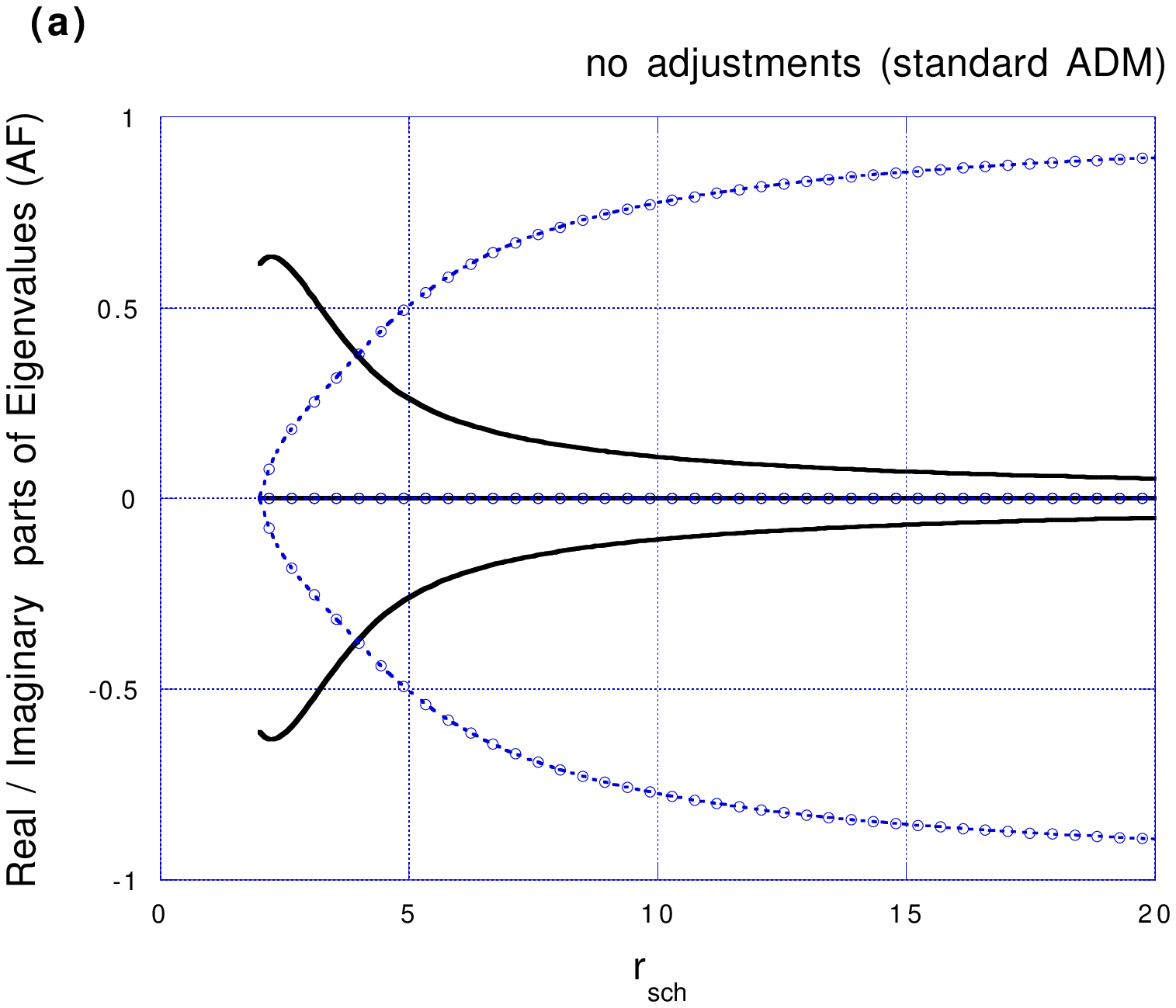} }
\put(1.0,0){\epsfxsize=6.0cm \epsffile{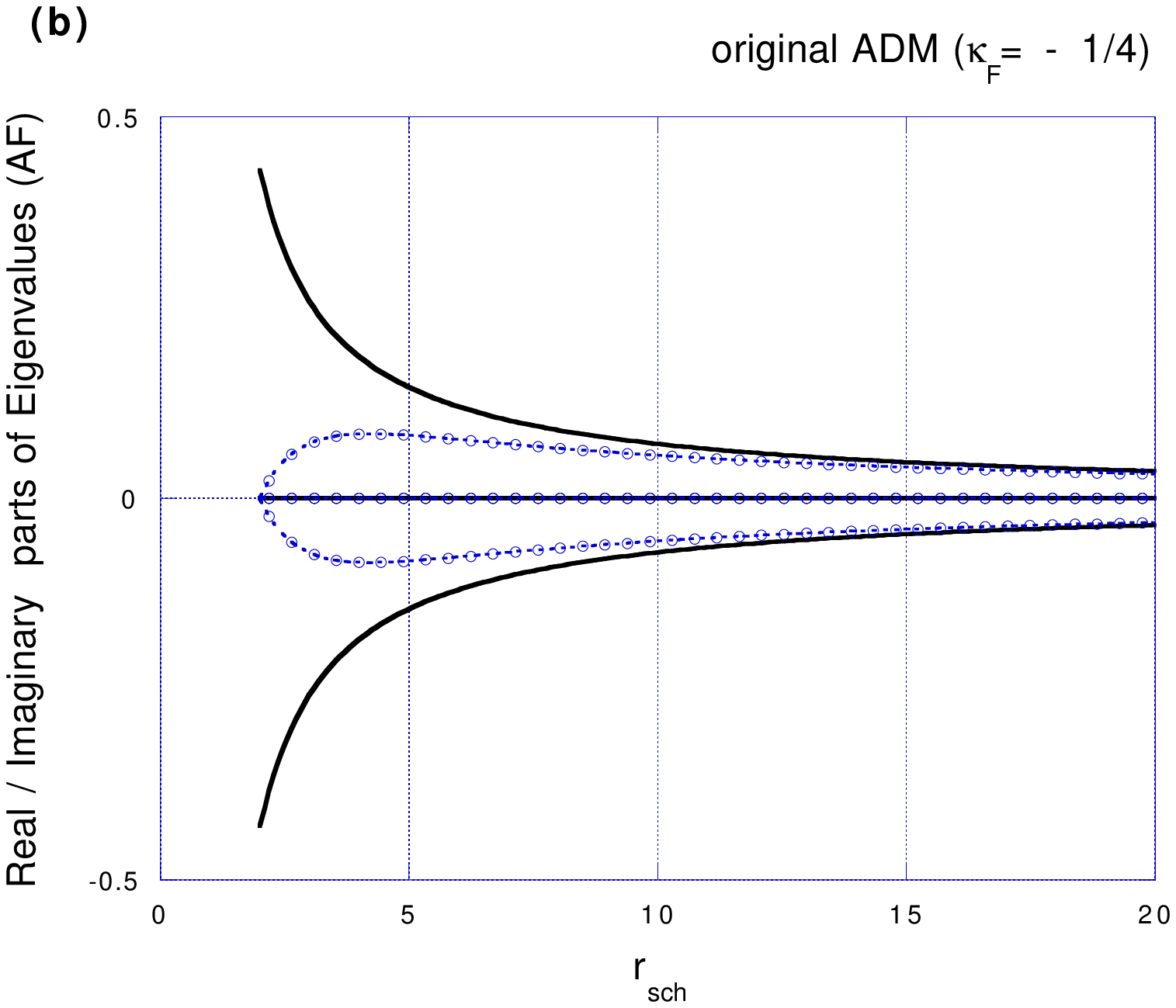} }
\put(8.5,0){\epsfxsize=6.0cm \epsffile{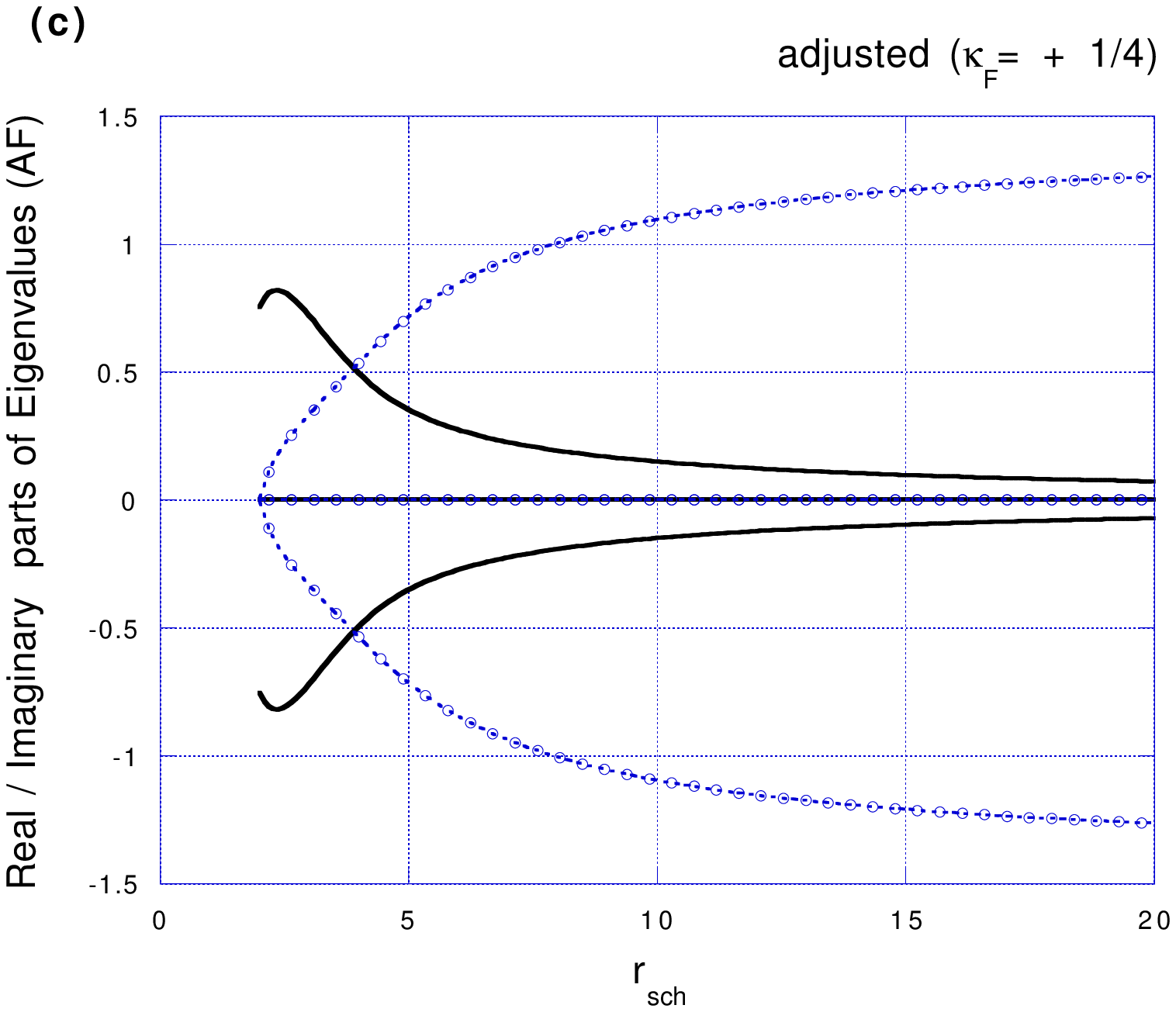} }
\end{picture}
\caption[quartic]{
Amplification factors (AFs, eigenvalues of 
homogenized constraint propagation 
equations) 
are shown for 
the standard Schwarzschild coordinate, with 
(a) no adjustments, i.e., standard ADM, 
(b) original ADM ($\kappa_F=-1/4$) and 
(c) an adjusted version with different signature, $\kappa_F=+1/4$ 
[see eq. (\ref{originalADMadjust})]. 
The solid lines and the dotted lines with circles are 
real parts and imaginary parts, respectively.  
They are four lines each, but 
actually the two eigenvalues are zero for all cases.  
Plotting range is $2 < r \le 20$ using
Schwarzschild radial coordinate. We set $k=1, l=2,$ and 
$m=2$ throughout the article. 
}
\label{fig1schadm}
\end{figure}

In Fig.\ref{fig1schadm}(a), the amplification factors (AFs, the 
eigenvalues of homogenized 
constraint propagation equations) of the standard ADM formulation
are plotted.   
The solid lines and dotted lines with circles are 
real parts and imaginary parts of AFs, respectively.   
(The figure style is common throughout the article.)
They are four lines each, but 
as we showed in (\ref{eigenvalueADM}), two of them are zero. 
The plotting range is $2 < r \le 20$ in Schwarzschild 
radial coordinate. The
AFs at $r=2$ are $\pm \sqrt{3/8}$ and $0$. 
The existence of this positive real AF near the horizon
is a new result which was not
seen in the flat background \cite{ronbun3}. 
We show only the cases with $l=2$ and $k=1$, because we judged that the plots of 
$l=0$ and other $k$s are qualitatively the same. 

The adjustment (\ref{originalADMadjust}) with 
$\kappa_F=-1/4$ returns the system 
back to the original ADM.  AFs are (\ref{eigenvalueOADM}) 
and we plot them in 
Fig.\ref{fig1schadm}(b).  We can see that the imaginary parts are 
apparently different from those of 
the standard ADM [Fig.\ref{fig1schadm}(a)].  
This is the same feature 
as in the case of the flat background \cite{ronbun3}.
According to our conjecture, 
the non-zero imaginary values are better than zeros, so 
we expect that the standard
ADM has a better stability than the original ADM system.  
Negative  $\kappa_F$ makes the asymptotical 
real values finite. 
If we change the signature of $\kappa_F$, then AFs are as in 
Fig.\ref{fig1schadm}(c).
The imaginary parts become larger than $\kappa_F=0$, that 
indicates the constraint
propagation involves higher frequency modes. 

\begin{figure}[p]
\setlength{\unitlength}{1cm}
\begin{picture}(15,11)
\put(1.0,5.5){\epsfxsize=6.0cm \epsffile{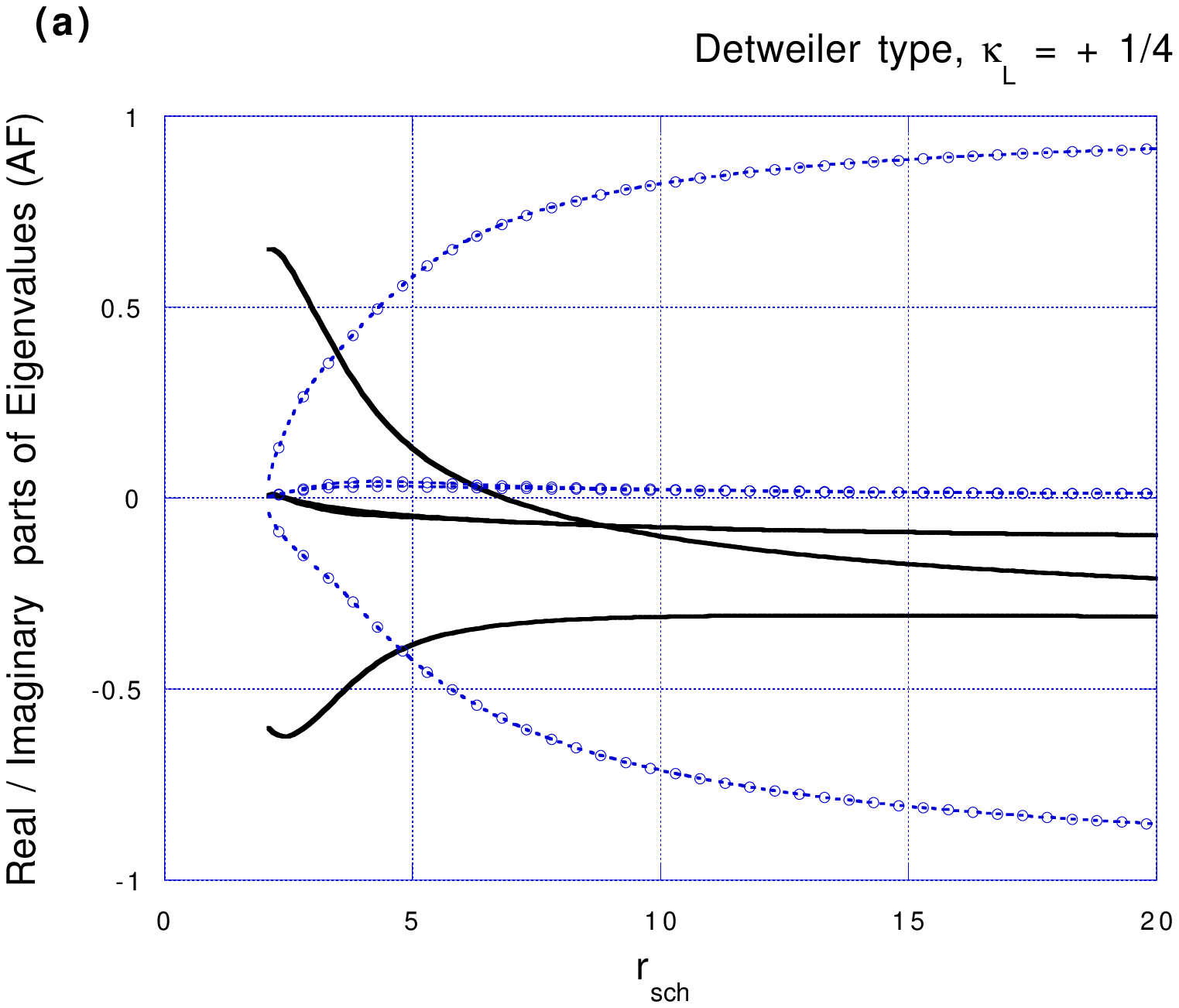} }
\put(8.5,5.5){\epsfxsize=6.0cm \epsffile{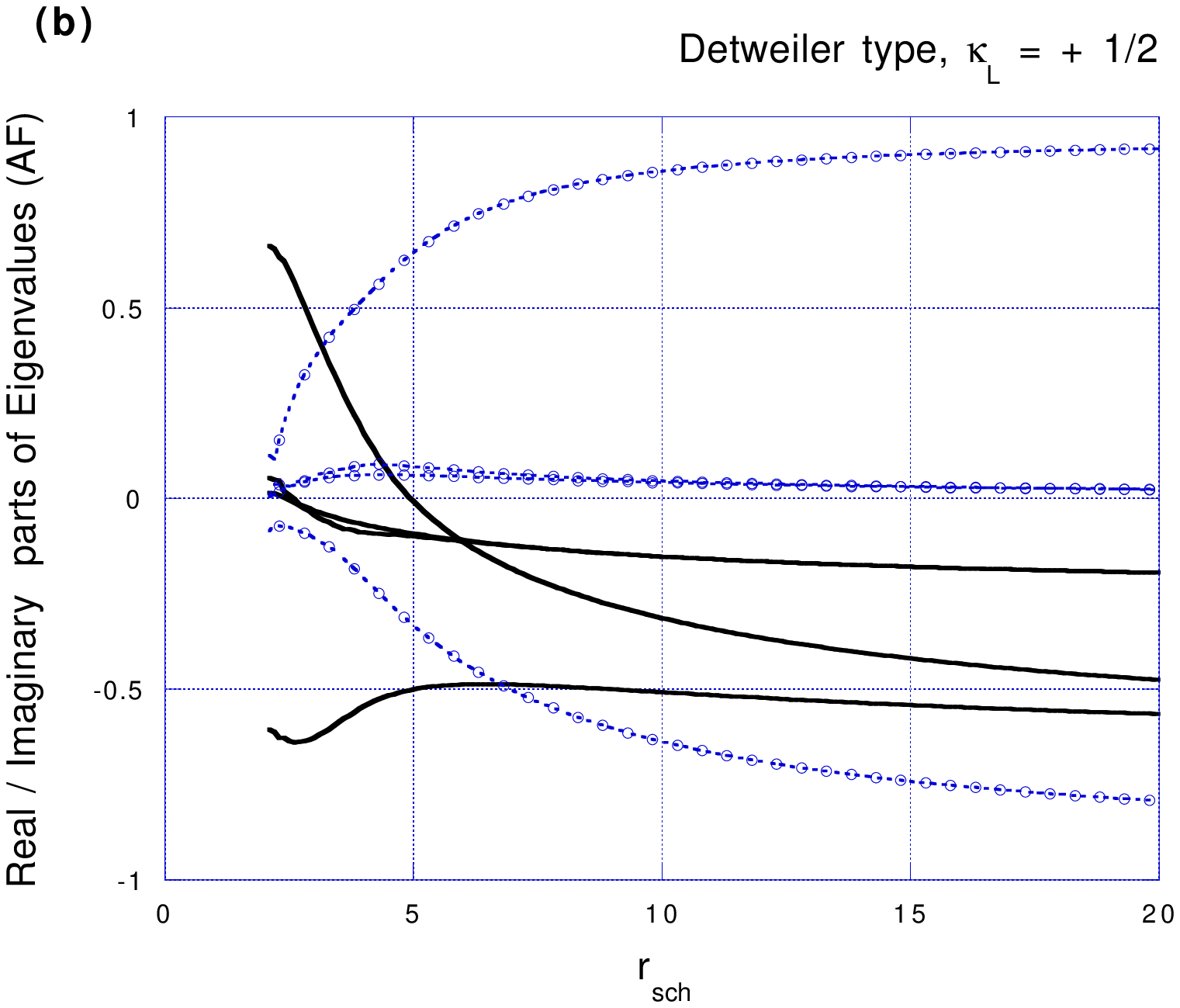} }
\put(1.0,0)  {\epsfxsize=6.0cm \epsffile{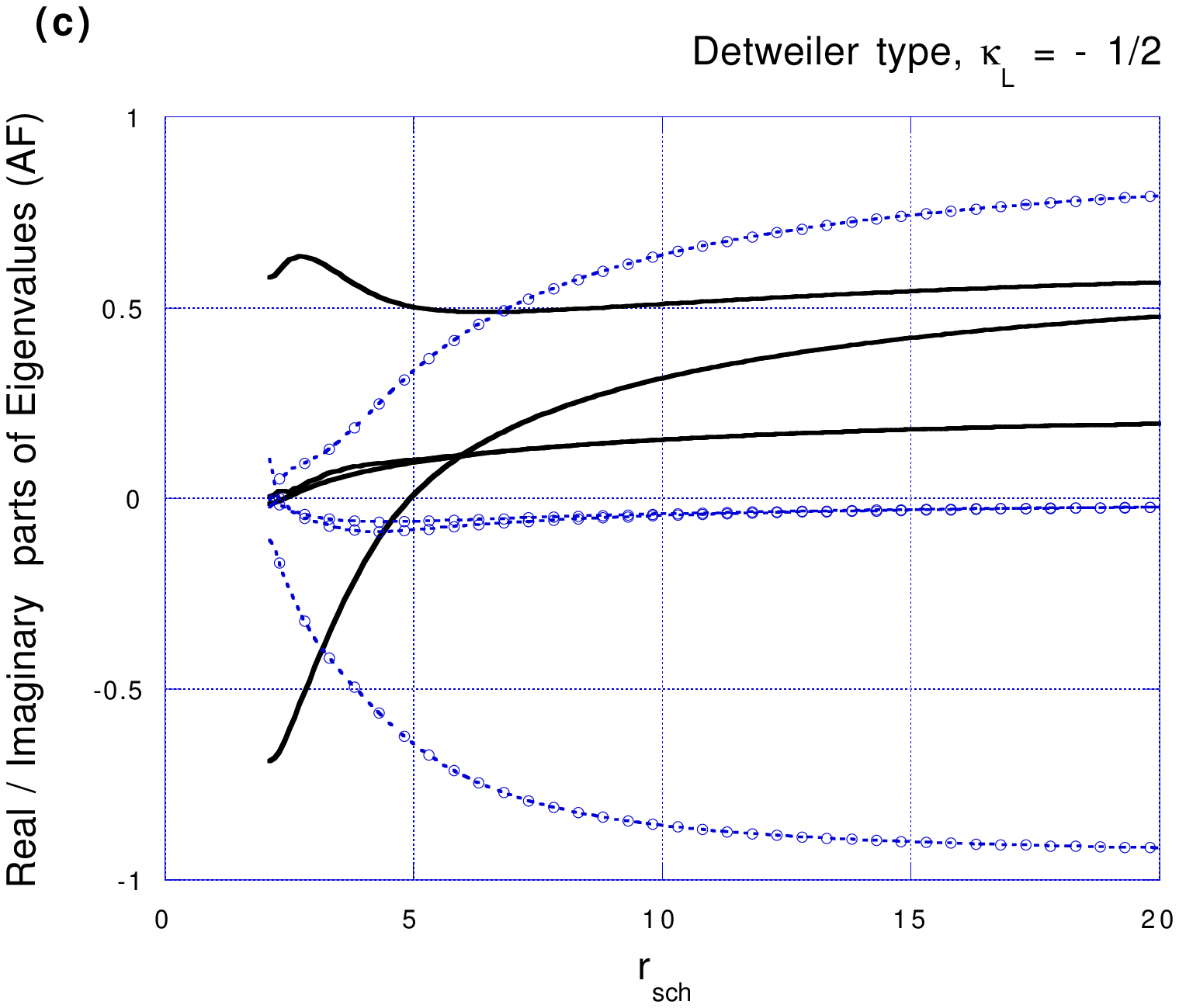} }
\end{picture}
\caption[quartic]{
Amplification factors of the standard Schwarzschild coordinate, with 
Detweiler type adjustments, (\ref{Det1})-(\ref{Det4}). 
Multipliers used in the plot are  (a) $\kappa_L=+1/4$, 
(b) $\kappa_L=+1/2$, and 
(c) $\kappa_L=-1/2$.
Plotting details are the same as Fig.\ref{fig1schadm}. 
}
\label{fig2schdet}
\end{figure}

The adjustment proposed by Detweiler, (\ref{Det1})-(\ref{Det4}), 
makes the feature 
completely different.  
Fig.\ref{fig2schdet}(a) and (b) are the case of $\kappa_L=1/4$
and $1/2$ and (c) is of the different signature of $\kappa_L$.  
A great improvement can be seen in both positive $\kappa_L$ cases 
where {\it all} real parts 
become negative in large $r$. 
Moreover all imaginary parts are apart from zero. 
These are 
the desired features according to our conjecture. 
Therefore we expect the Detweiler adjustment has good stability properties
{\it except} near the black hole region. 
The AF near the horizon {\it has} a positive real component. 
This is not contradictory with the 
Detweiler's original idea.  The idea  
 came from 
suppressing the {\it total} L2 norm of constraints on the spatial slice, 
while our plot indicates 
the existence of a {\it local} violation mode.   
The change of signature of $\kappa_L$
can be understood just by changing the signature of AFs, and 
this fact can also be seen  to the other examples.

\begin{figure}[p]
\setlength{\unitlength}{1cm}
\begin{picture}(15,11)
\put(1.0,5.5){\epsfxsize=6.0cm \epsffile{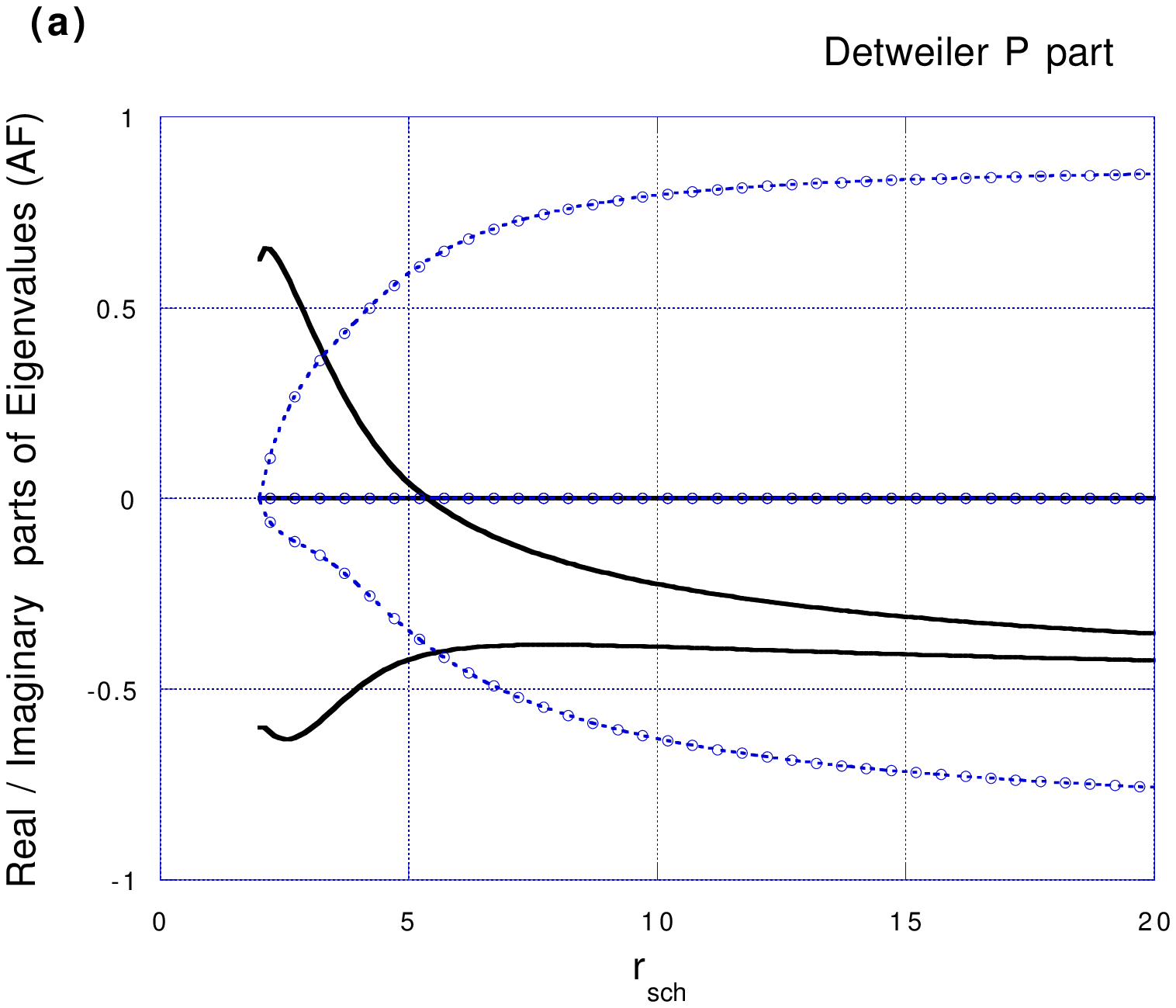} }
\put(8.5,5.5){\epsfxsize=6.0cm \epsffile{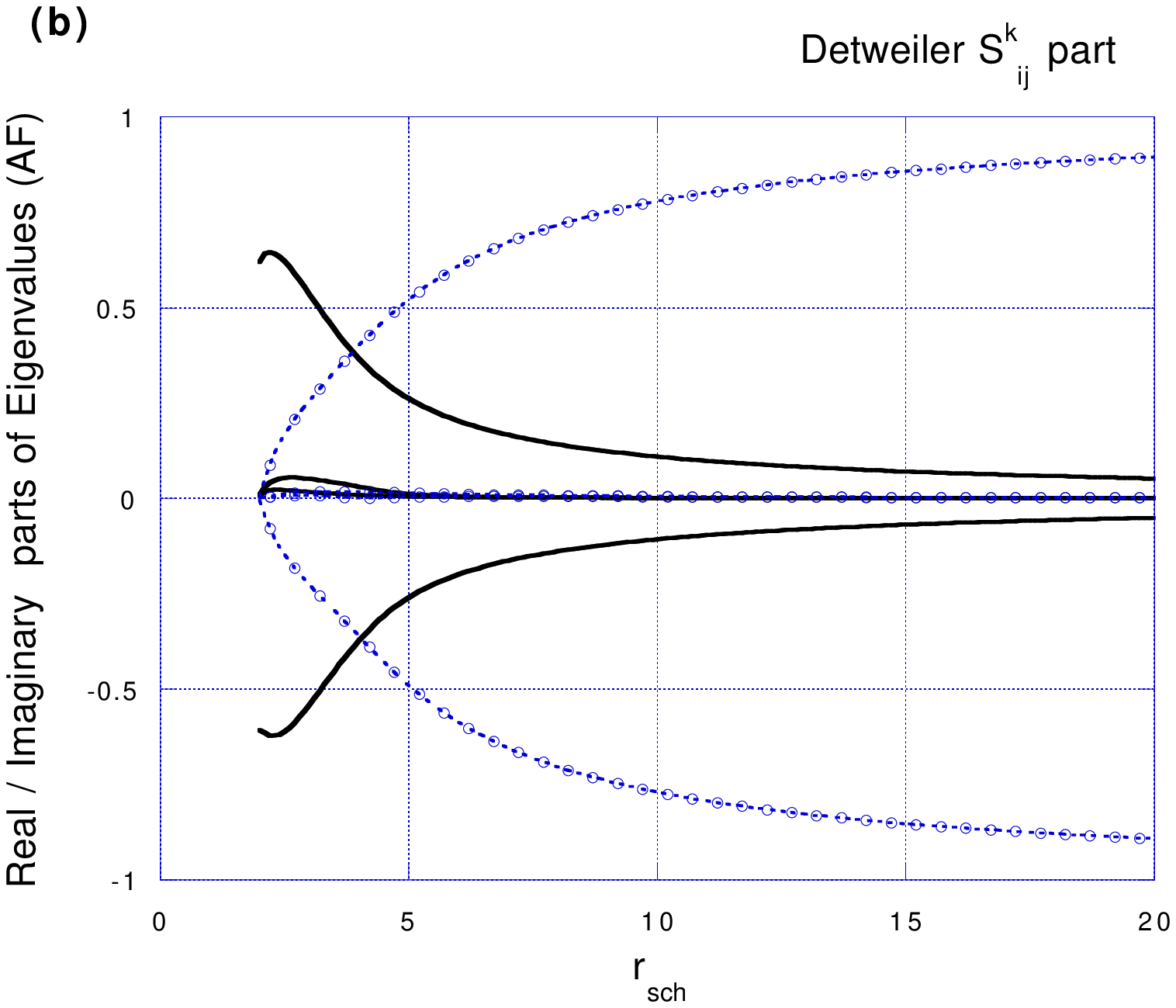} }
\put(1.0,0.0){\epsfxsize=6.0cm \epsffile{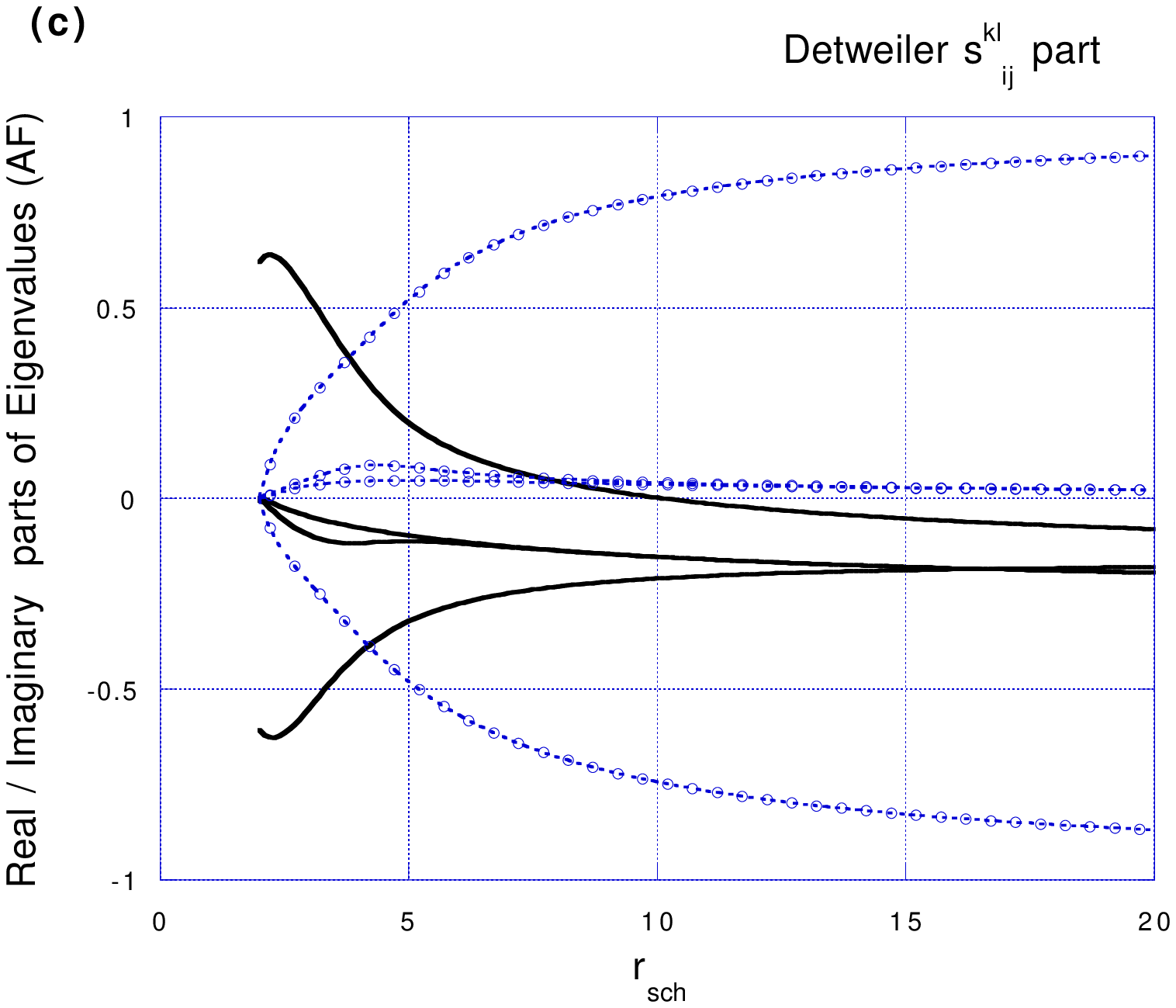} }
\put(8.5,0.0){\epsfxsize=6.0cm \epsffile{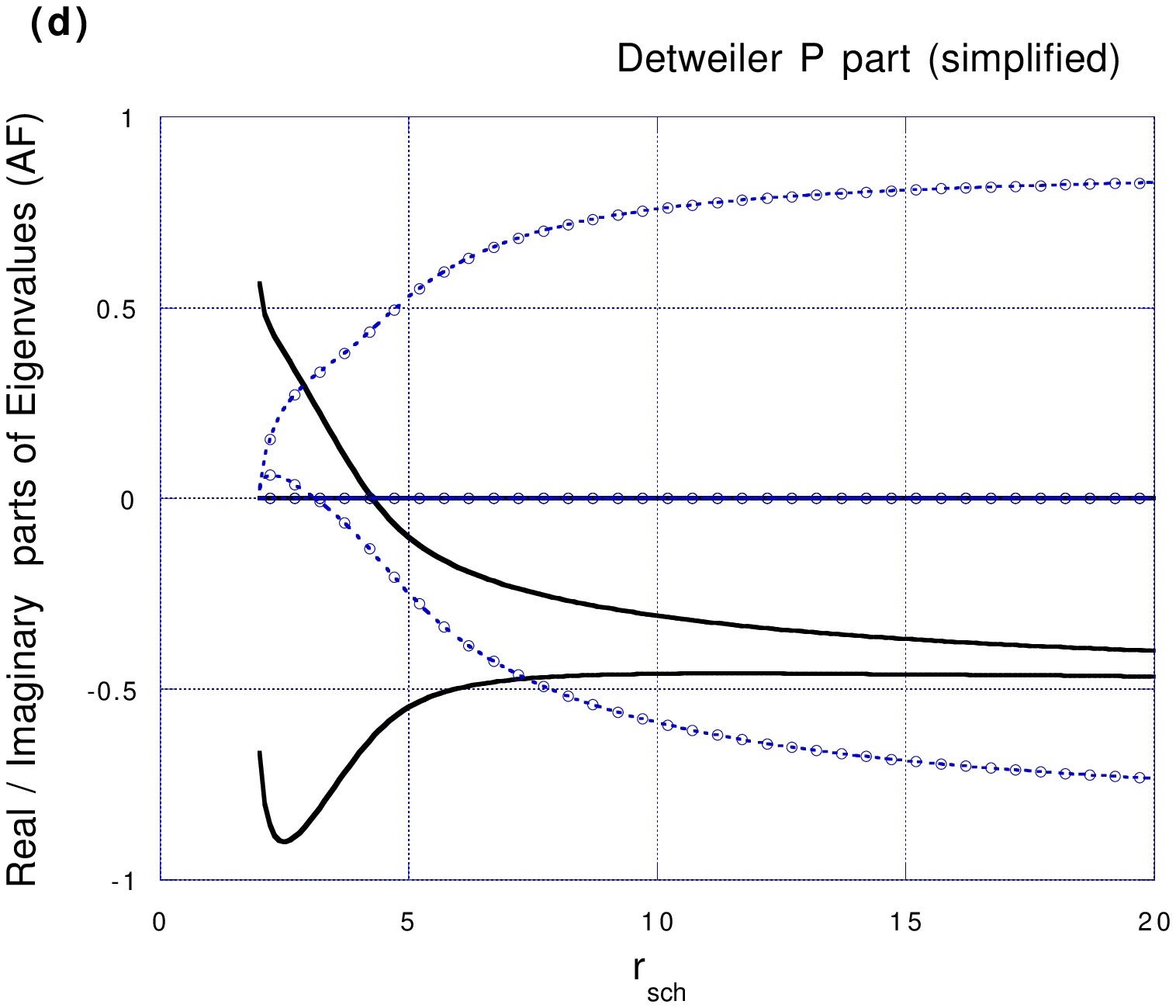} }
\end{picture}
\caption[quartic]{ 
Amplification factors of the standard Schwarzschild coordinate, with 
Detweiler type (partial contributions) adjustments. 
Fig.(a) is the case No. 2-P in Table.\ref{table1}, 
adjusting with $P_{ij}=-\kappa_L \alpha^3 \gamma_{ij}$, i.e. (\ref{Det1}), 
and else zero. 
Similarly, 
(b) is No. 2-S,  $S^k{}_{ij}$-part (\ref{Det3})  and else zero,  and
(c) is No. 2-s,   $s^{kl}{}_{ij}$-part  (\ref{Det4})  and else zero. 
Fig. (d) is No. 3,  $P_{ij}=-\kappa_L \alpha \gamma_{ij}$ and else zero, 
which is a  minor modification from (a).
 We used $\kappa_L=+1/2$ for all plots. 
Plotting details are the same as Fig.\ref{fig1schadm}. 
}
\label{fig3schdetpart}
\end{figure}

We next show that 
the partial adjustments of Detweiler's
are also not so bad.  
Fig.\ref{fig3schdetpart} (a), (b) and (c) are the cases of 
adjustment that are of only 
(\ref{Det1}), (\ref{Det3}) and (\ref{Det4}), respectively.  
[The contribution of 
(\ref{Det2}) is absent since $K_{ij}=0$ in the Schwarzschild coordinate.]
By comparing them with 
Fig.\ref{fig1schadm}(a) and Fig.\ref{fig2schdet}(b), we see the 
negative real parts in large $r$ are originated by 
(\ref{Det1}) and (\ref{Det4}), while the former two real parts 
remain zero. 
Fig.\ref{fig3schdetpart}(d) used only 
$P_{ij}=-\kappa_L \alpha \gamma_{ij}$ and everything else is zero, 
which is a minor modification from (\ref{Det1}).  
The contribution is similar, and can be said to be {\it effective}. 

\subsection{in isotropic/iEF/PG coordinates}

\begin{figure}[p]
\setlength{\unitlength}{1cm}
\begin{picture}(15,11)
\put(1.0,5.5){\epsfxsize=6.0cm \epsffile{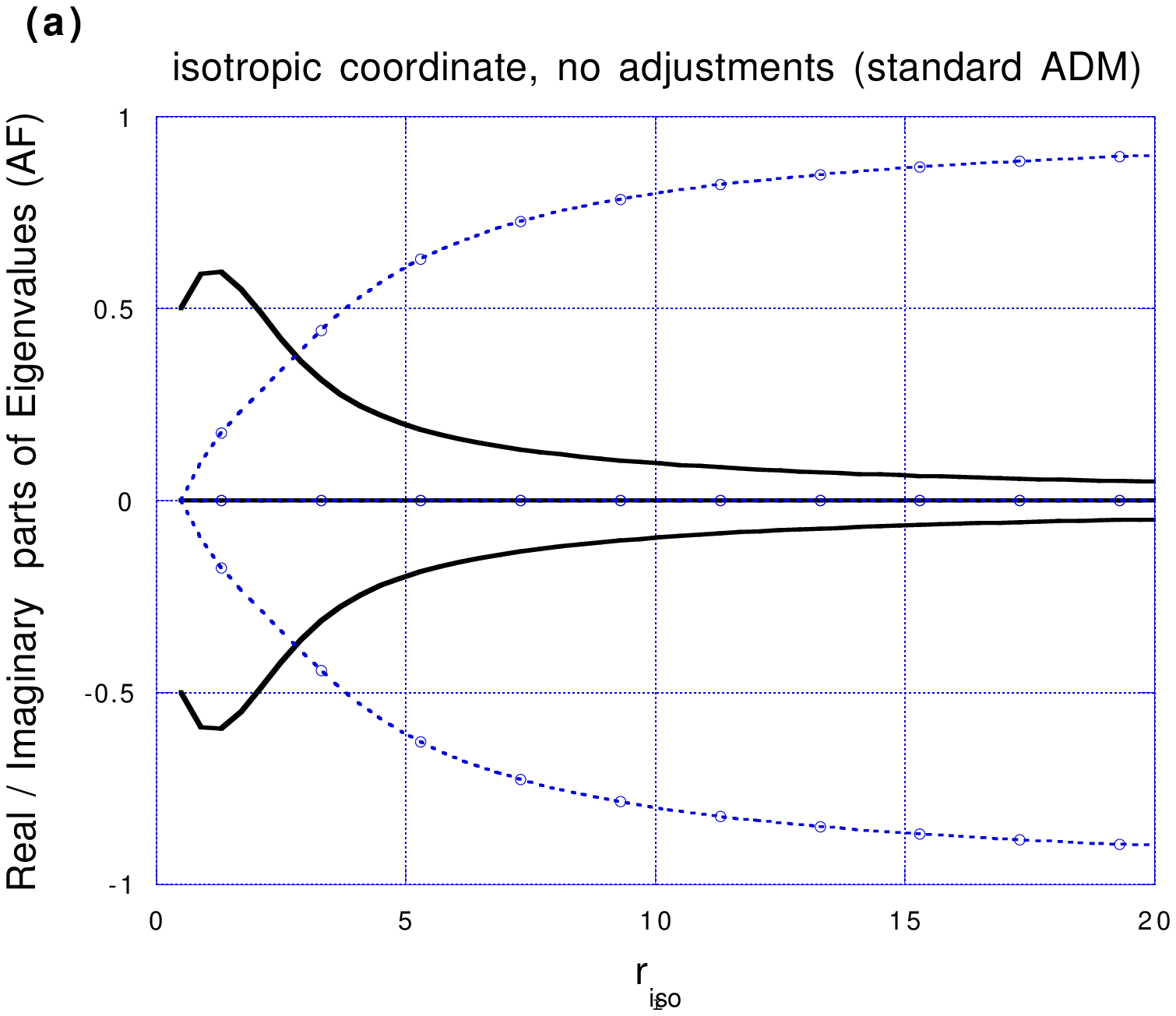} }
\put(1.0,0){\epsfxsize=6.0cm \epsffile{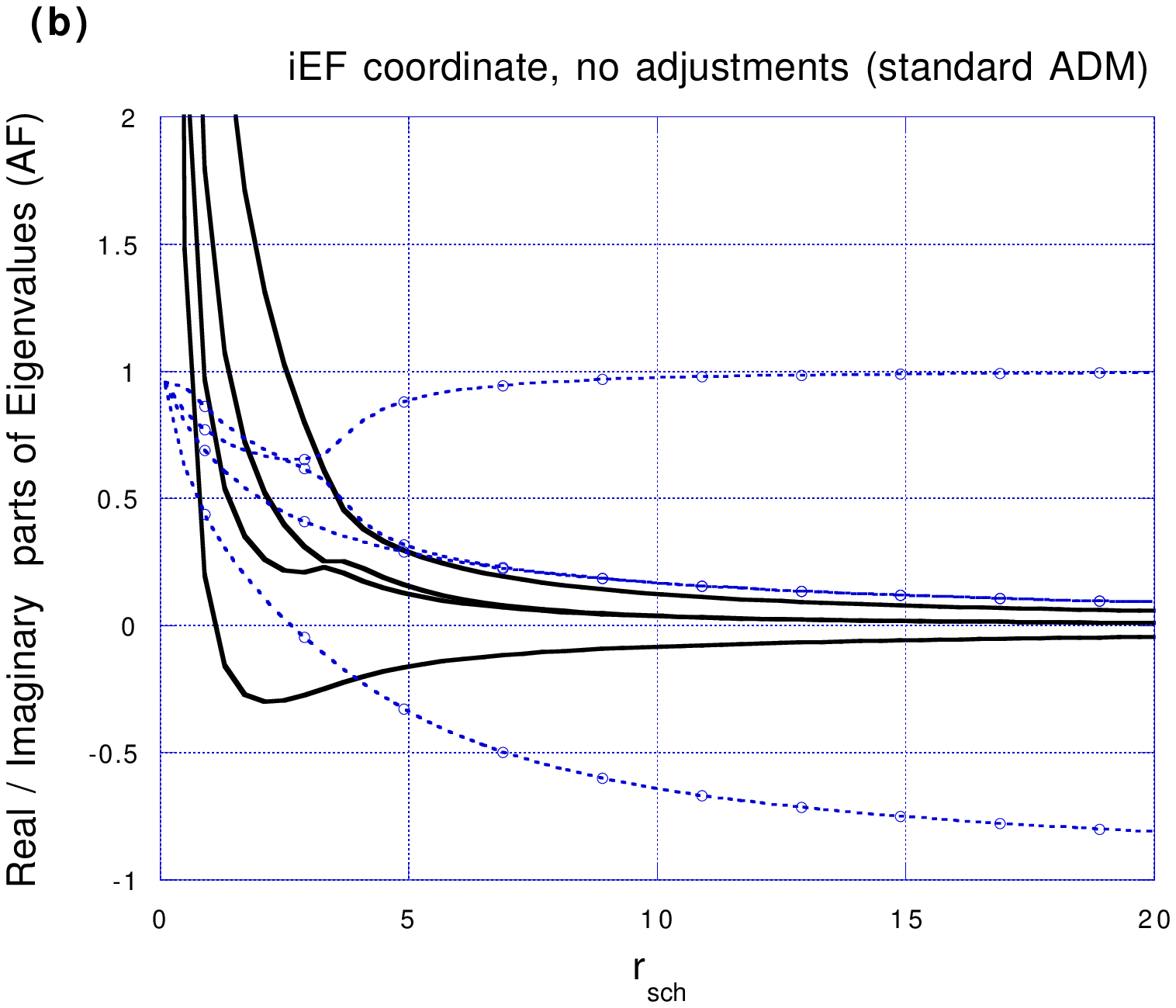} }
\put(8.5,0){\epsfxsize=6.0cm \epsffile{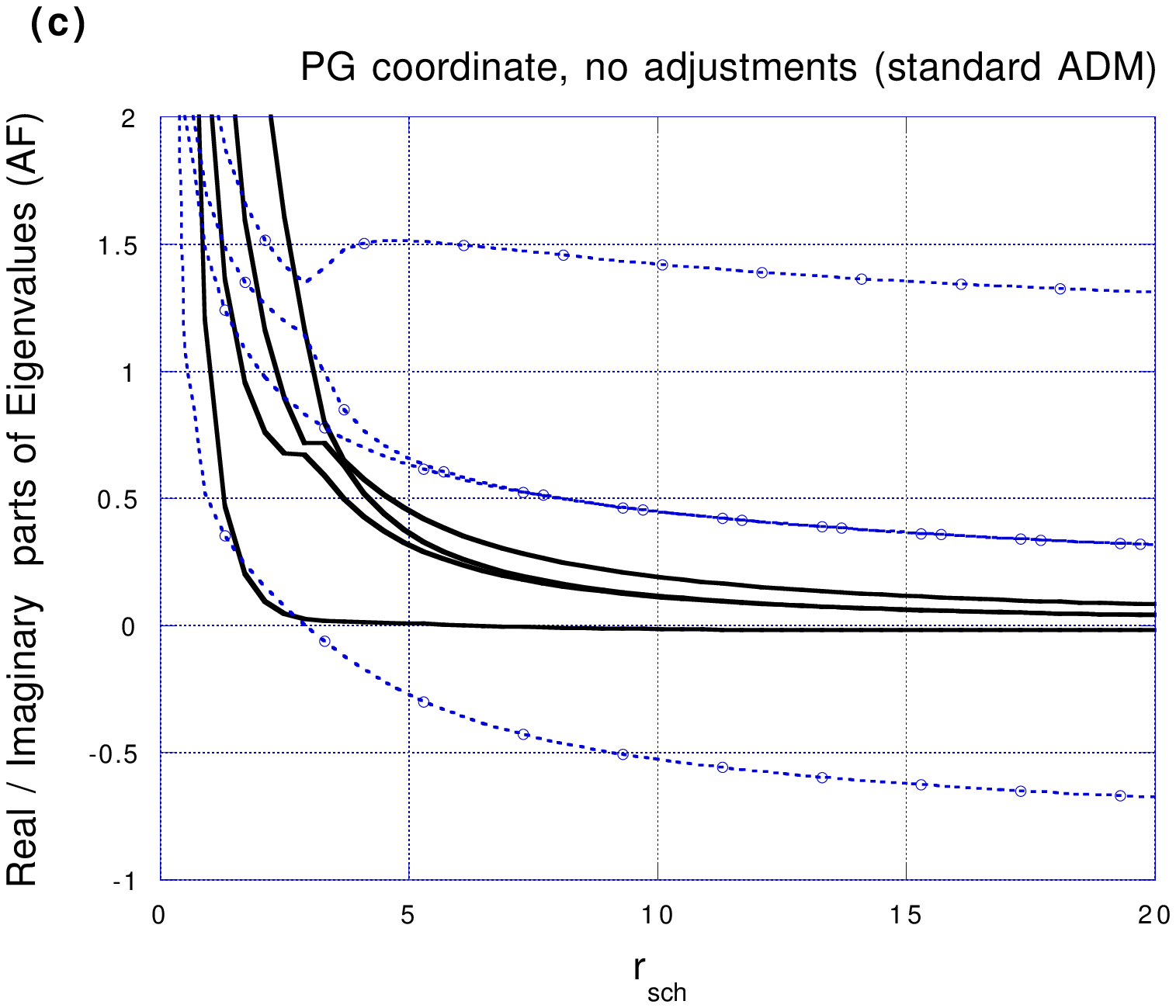} }
\end{picture}
\caption[quartic]{
Comparison of 
amplification factors between different coordinate expressions
for  the standard ADM formulation (i.e. no adjustments). 
Fig. (a) is for the isotropic coordinate (\ref{isoSch}), and the 
plotting range is $1/2 \le r_{iso}$. 
Fig. (b) and (c) are for the iEF coordinate (\ref{iEFSch}) and
the PG coordinate (\ref{PGSch}), respectively, and we plot
lines on the $t=0$ slice for each expression. 
[See Fig.\ref{fig1schadm}(a) for the standard Schwarzschild coordinate.]
The solid four lines and the dotted four lines with circles are 
real parts and imaginary parts, respectively.  
}
\label{fig4adm}
\end{figure}

We next compare AFs between different coordinate expressions.  
The first test is for the standard ADM formulation. 
Fig.\ref{fig4adm} shows AFs 
for (a) the isotropic coordinate (\ref{isoSch}), 
(b) the iEF coordinate (\ref{iEFSch}), and
(c) the PG coordinate (\ref{PGSch}).  All plots are on the time slice
of $t=0$ in each coordinate expression. 
[See Fig.\ref{fig1schadm}(a) for the standard Schwarzschild coordinate.]
We see that Fig.\ref{fig1schadm}(a) and Fig.\ref{fig4adm}(a) are quite
similar, while Fig.\ref{fig4adm}(b) and (c) are qualitatively different 
{}from the former.  This is because the latter expressions (iEF/PG) are 
asymmetric according to time, i.e. they have non-zero extrinsic curvature. 
  
We note that 
the constraint propagation
equations are invariant for spatial coordinate transformation, but
AFs are invariant only for linear transformation.  
This explains  the  
differences between Fig.\ref{fig1schadm}(a) and
Fig.\ref{fig4adm}(a), although these are not significant.  

\begin{figure}[p]
\setlength{\unitlength}{1cm}
\begin{picture}(15,11)
\put(1.0,5.5){\epsfxsize=6.0cm \epsffile{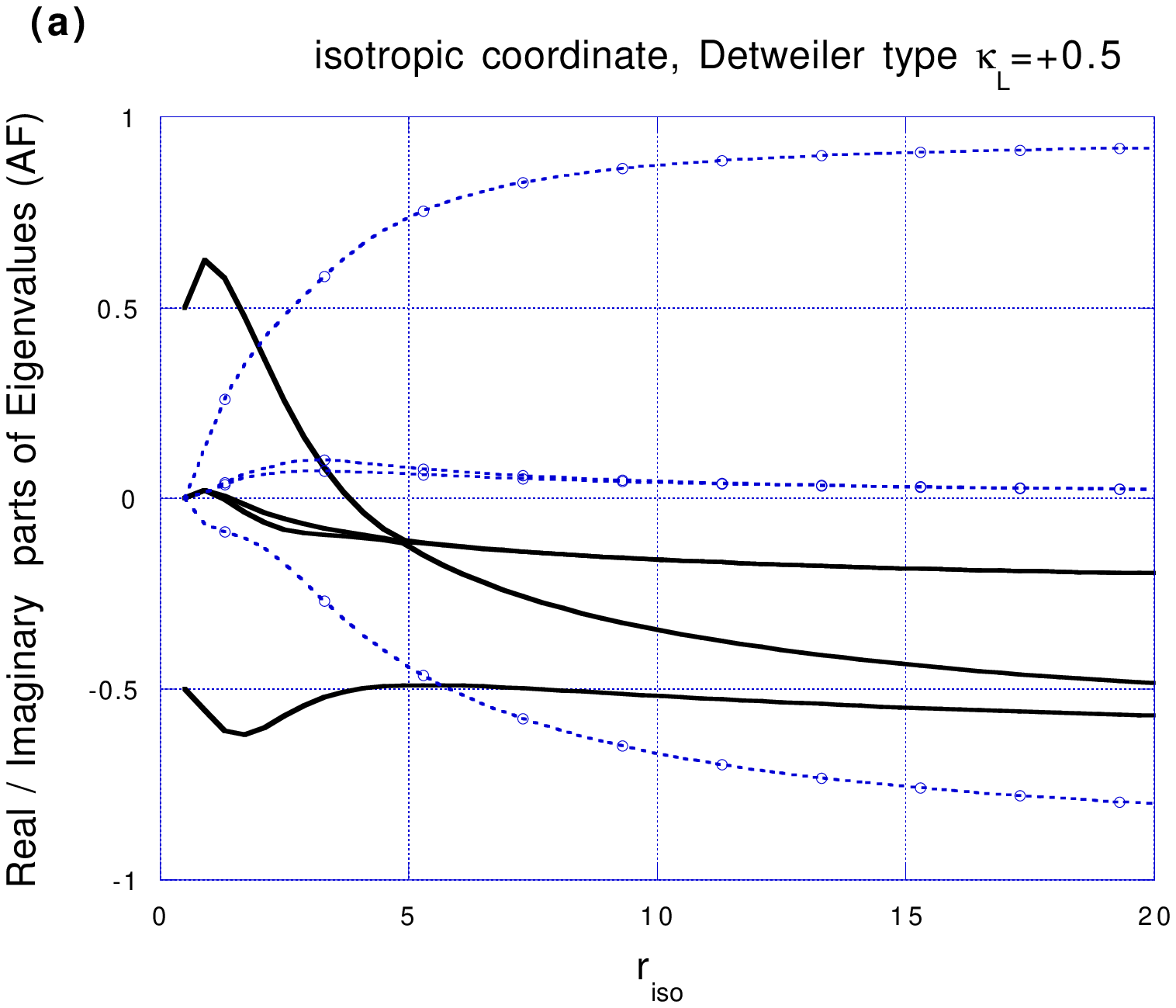} }
\put(1.0,0){\epsfxsize=6.0cm \epsffile{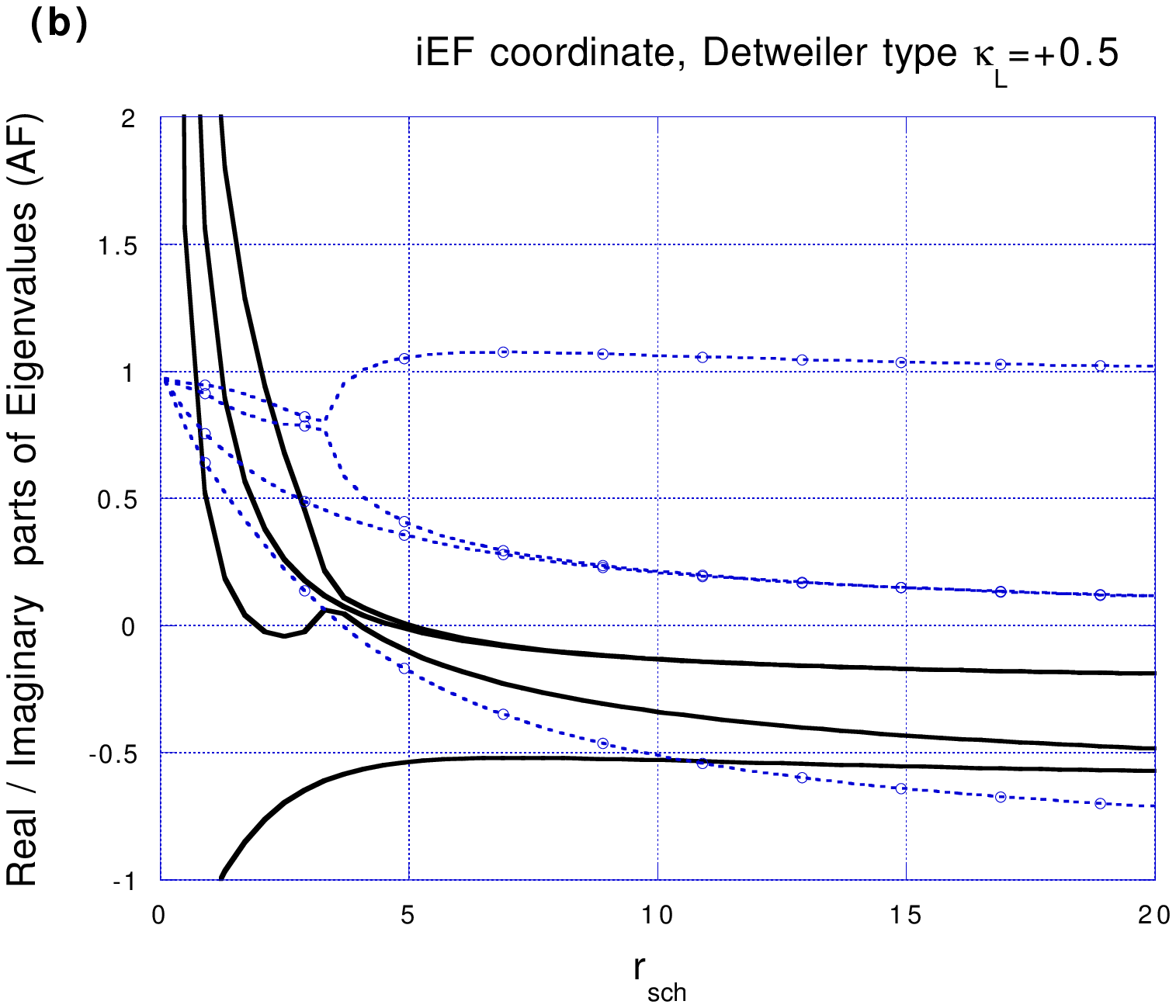} }
\put(8.5,0){\epsfxsize=6.0cm \epsffile{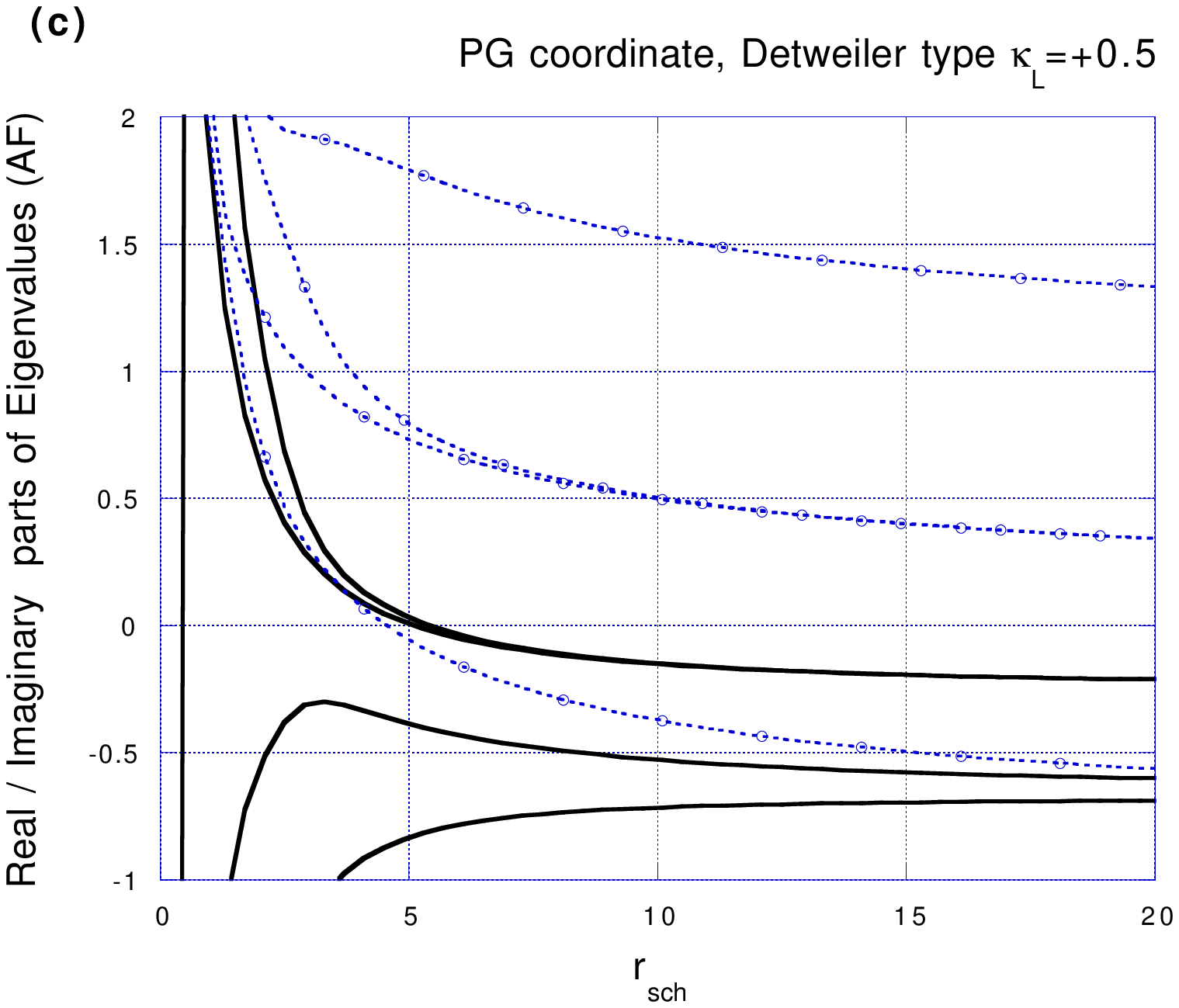} }
\end{picture}
\caption[quartic]{
Similar comparison with Fig.\ref{fig4adm}, but for 
Detweiler adjustments. $\kappa_L=+1/2$ for all plots.  
[See Fig.\ref{fig2schdet}(b) for the standard Schwarzschild 
coordinate.]
}
\label{fig5det}
\end{figure}

We notice that while some AFs in iEF/PG 
remain positive [Fig.\ref{fig4adm}(b) and (c)] in large $r$ region, 
that their nature changes due to the adjustments. 
Fig.\ref{fig5det} is for the Detweiler-type adjustment,
(\ref{Det1})-(\ref{Det4}),
for the isotropic/iEF/PG coordinate cases. 
[See Fig.\ref{fig2schdet}(b) for the standard Schwarzschild 
coordinate.]
Interestingly, all plots indicate that 
 all real parts of AFs are negative, and imaginary
parts are non-zero (again except near the  black hole region). 
By arranging the multiplier parameter, for the iEF/PG coordinates, 
there is a chance to get
all negative real AFs outside the black hole horizon. 
For example, for iEF (PG) coordinate all the real-part goes negative
outside the black hole horizon if $\kappa_L > 3.1 (1.6)$, while large
$\kappa_L$ may introduce another instability problem \cite{ronbun2}.   



\begin{figure}[p]
\setlength{\unitlength}{1cm}
\begin{picture}(15,11)
\put(1.0,5.5){\epsfxsize=6.0cm \epsffile{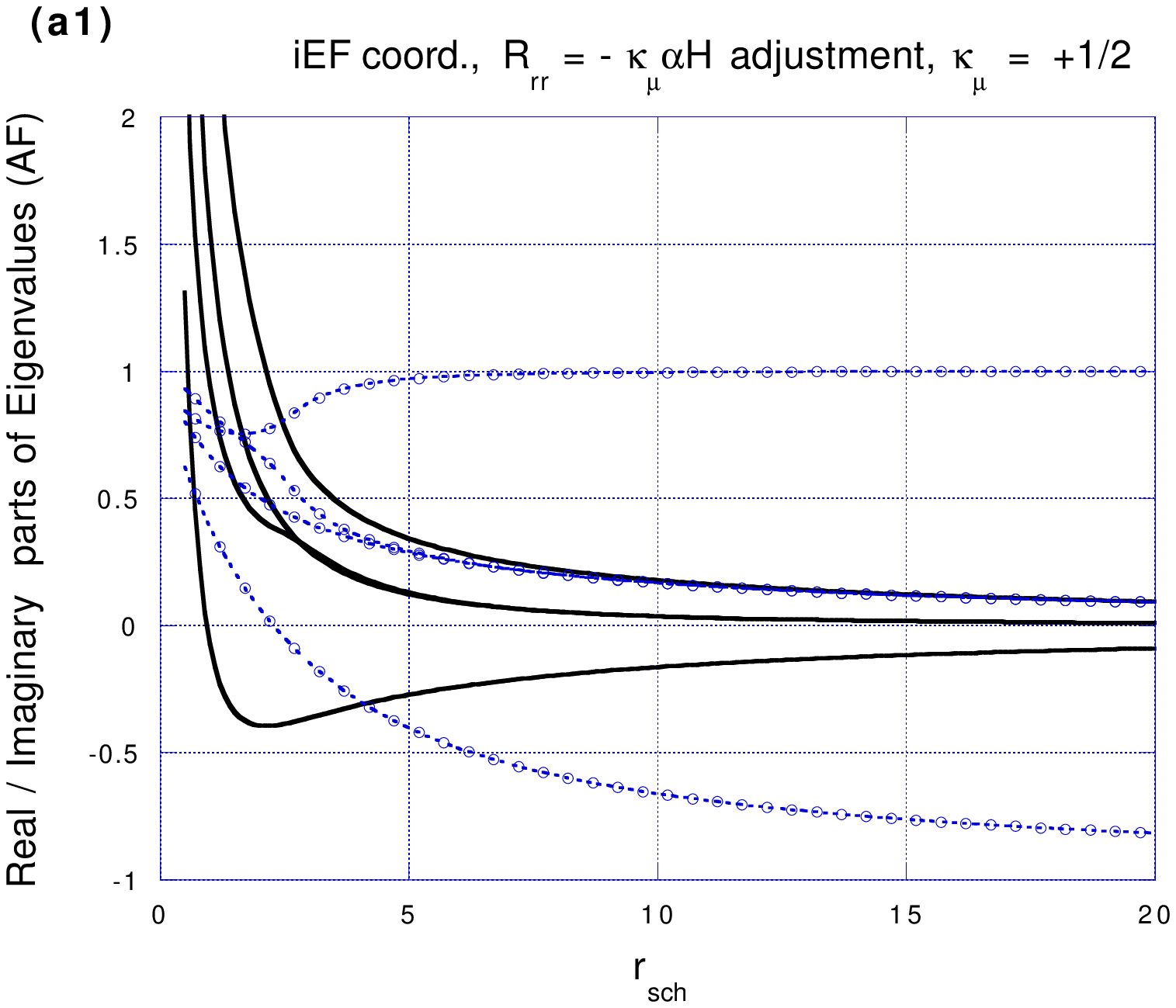} }
\put(8.5,5.5){\epsfxsize=6.0cm \epsffile{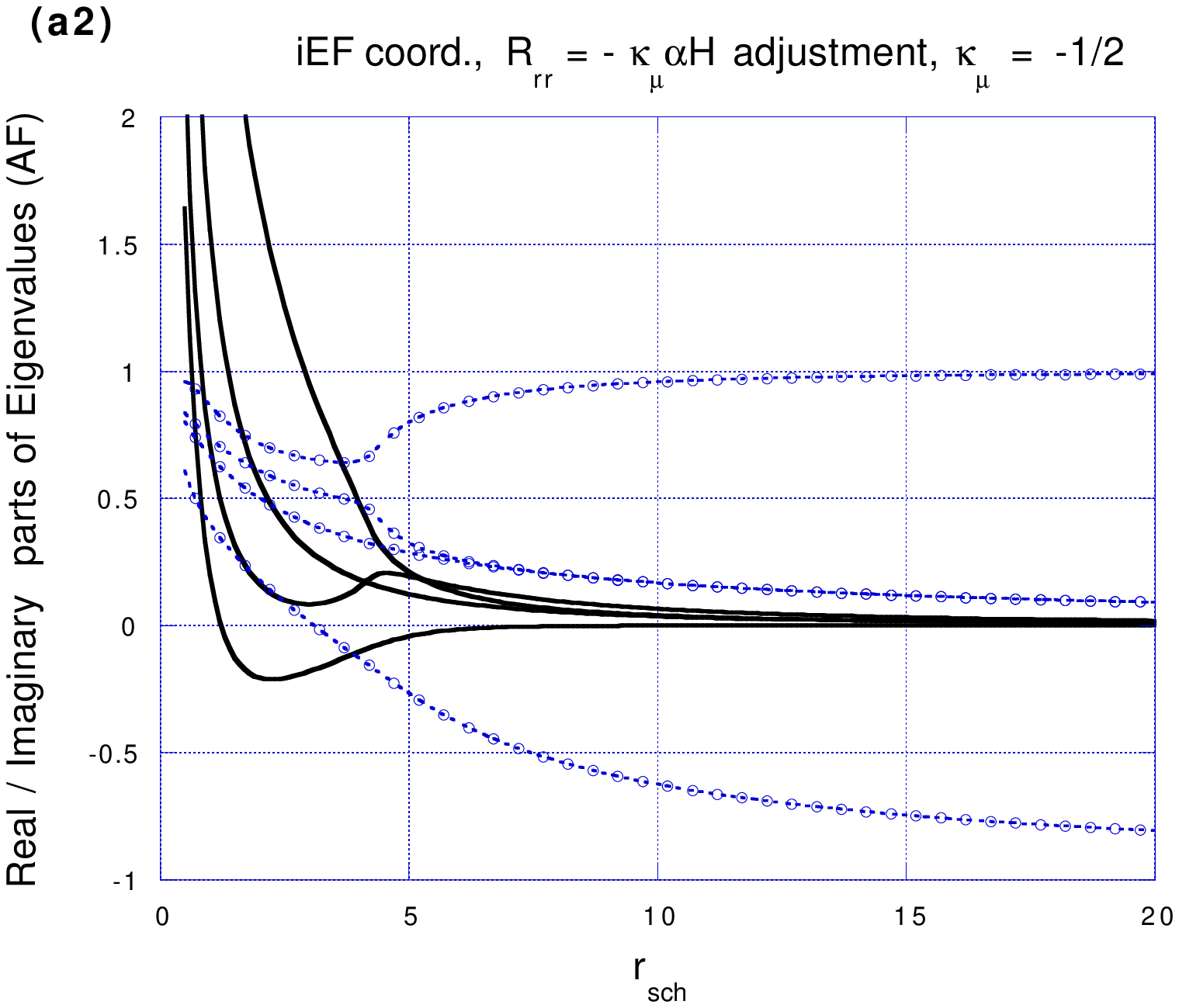} }
\put(1.0,0){\epsfxsize=6.0cm \epsffile{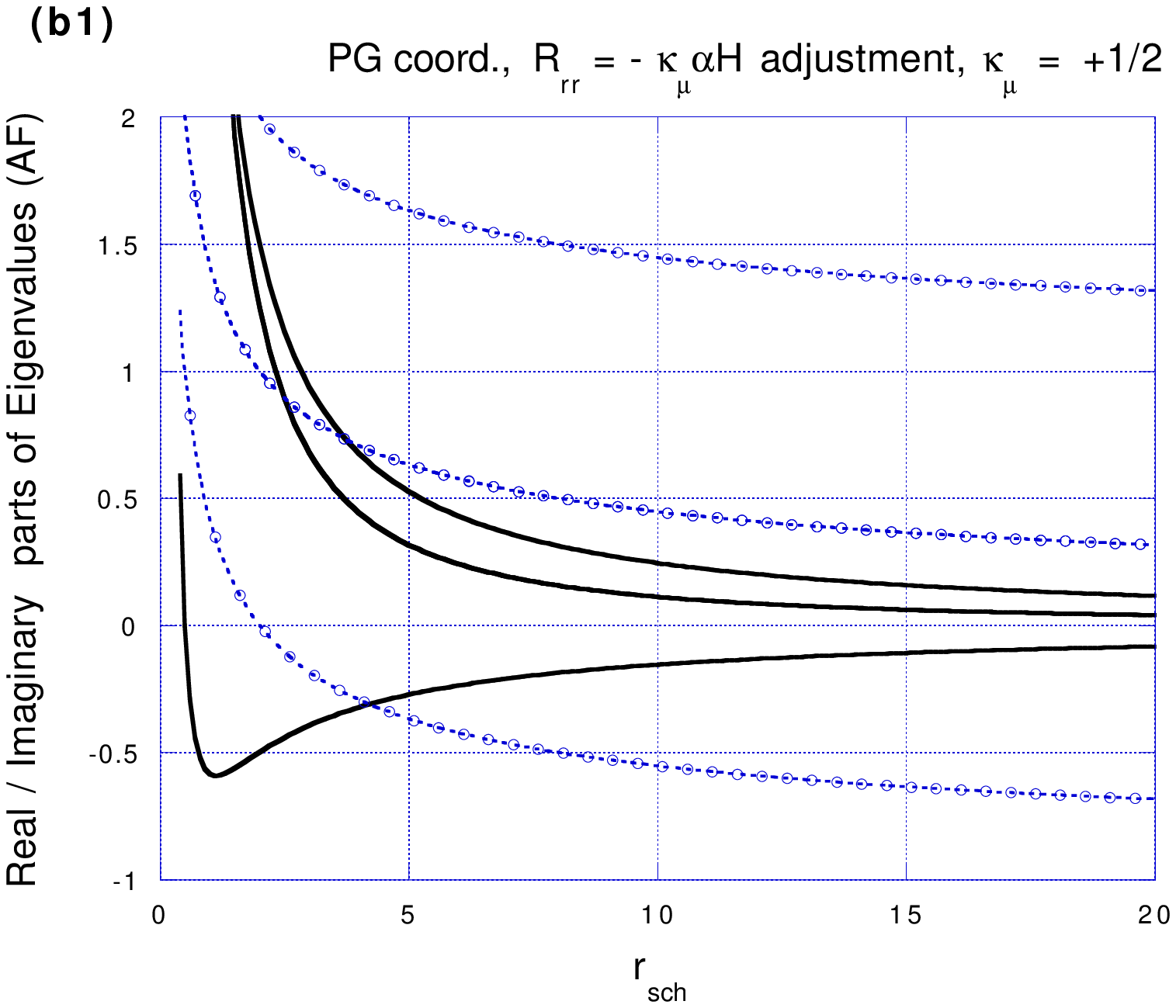} }
\put(8.5,0){\epsfxsize=6.0cm \epsffile{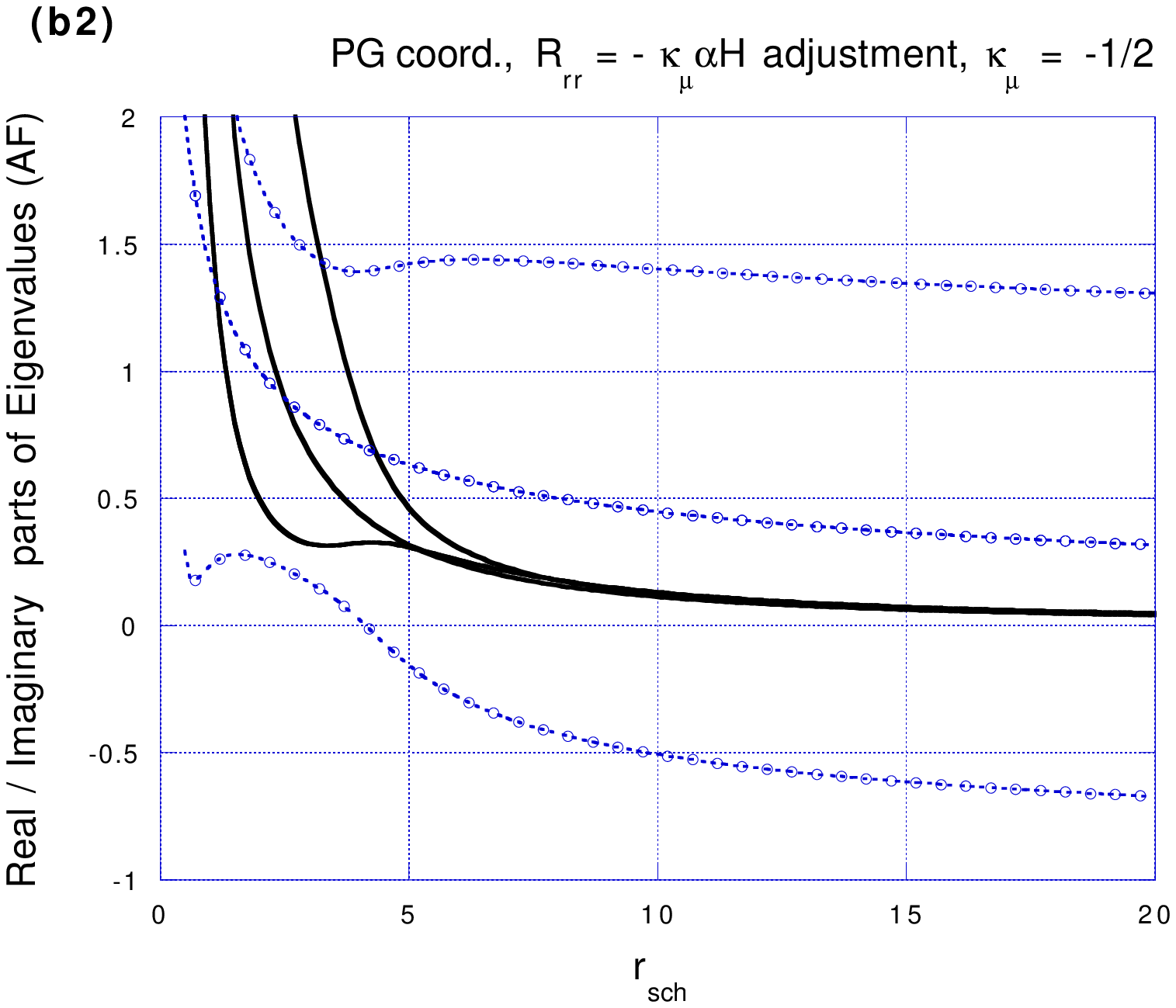} }
\end{picture}
\caption[quartic]{
Amplification factors of the adjustment  No.4 in Table. \ref{table1}. 
$\kappa_\mu=+1/2, -1/2$ for iEF/PG coordinates. 
[See Fig.\ref{fig4adm}(b)(c) for the standard ADM
system.]
}
\label{fig6penn}
\end{figure}

\begin{figure}[p]
\setlength{\unitlength}{1cm}
\begin{picture}(15,11)
\put(1.0,5.5){\epsfxsize=6.0cm \epsffile{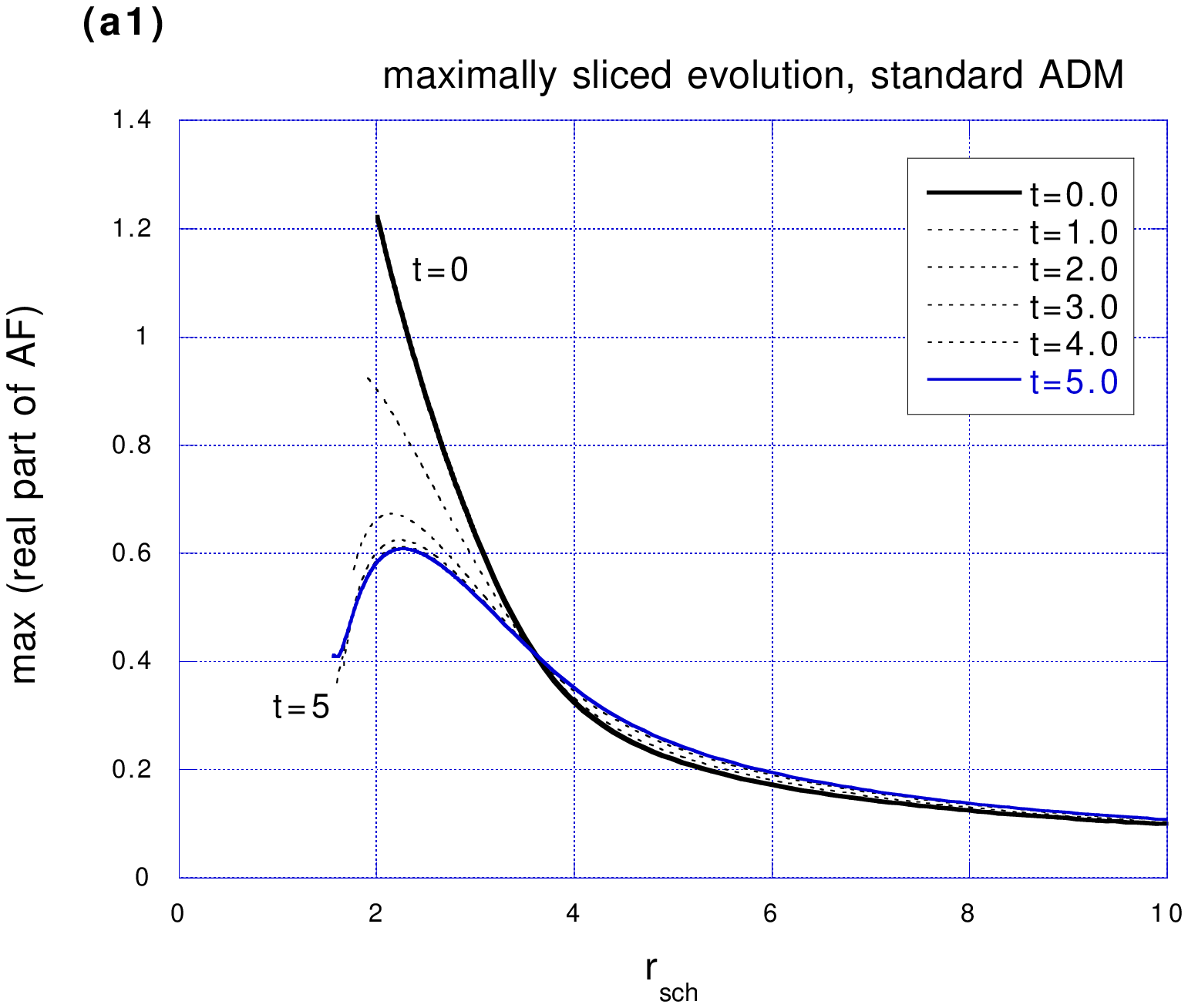} }
\put(8.5,5.5){\epsfxsize=6.0cm \epsffile{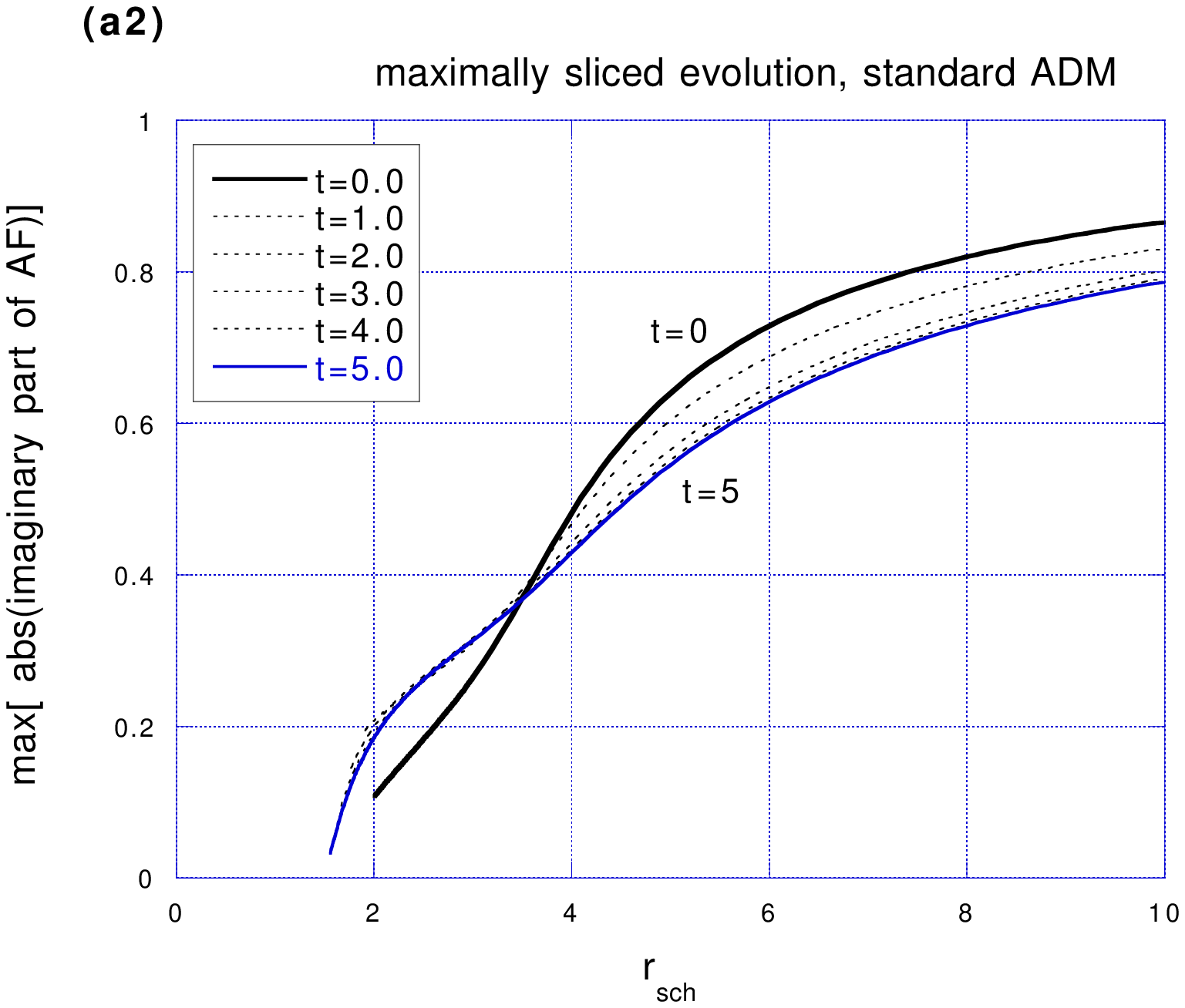} }
\put(1.0,0){\epsfxsize=6.0cm \epsffile{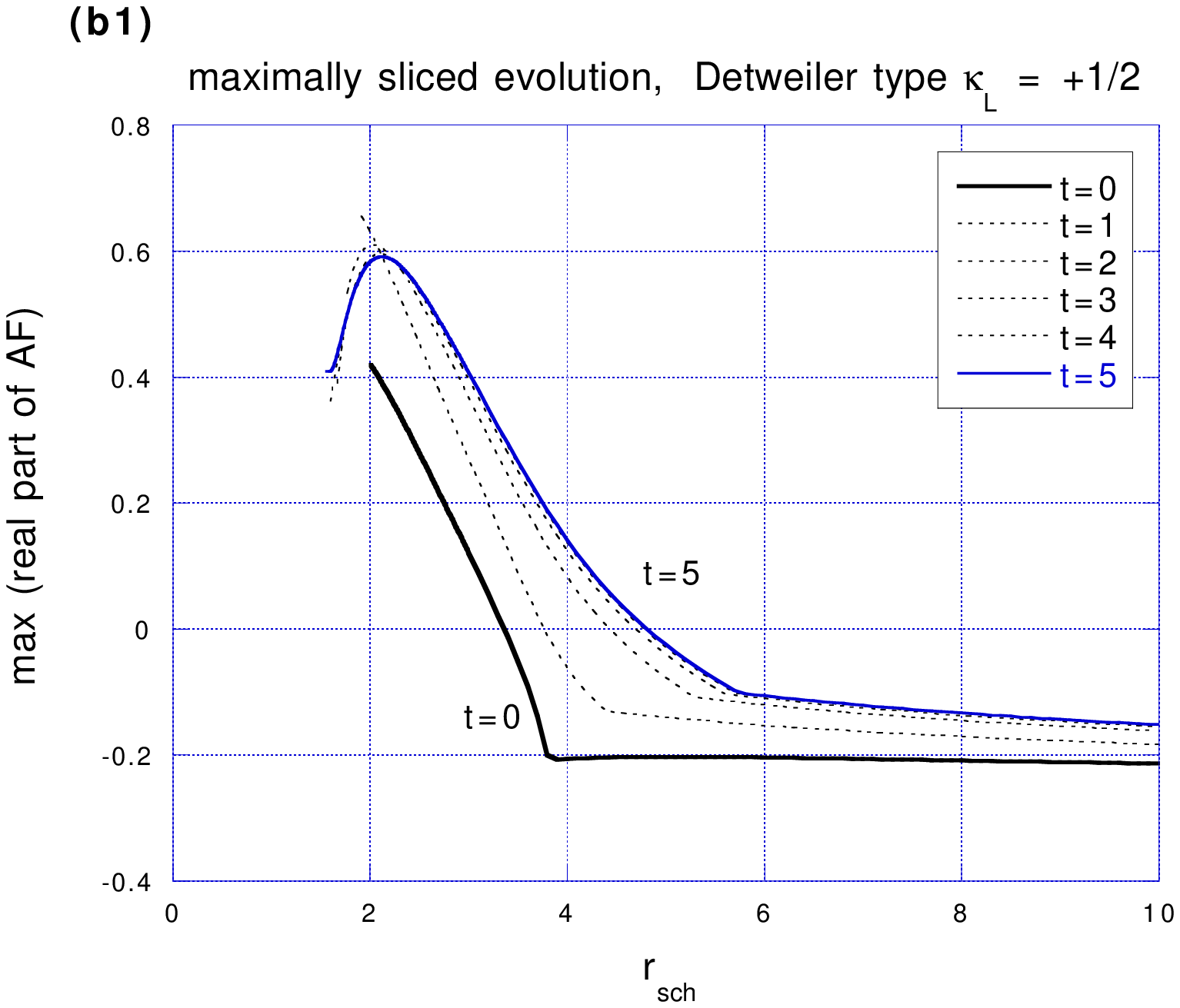} }
\put(8.5,0){\epsfxsize=6.0cm \epsffile{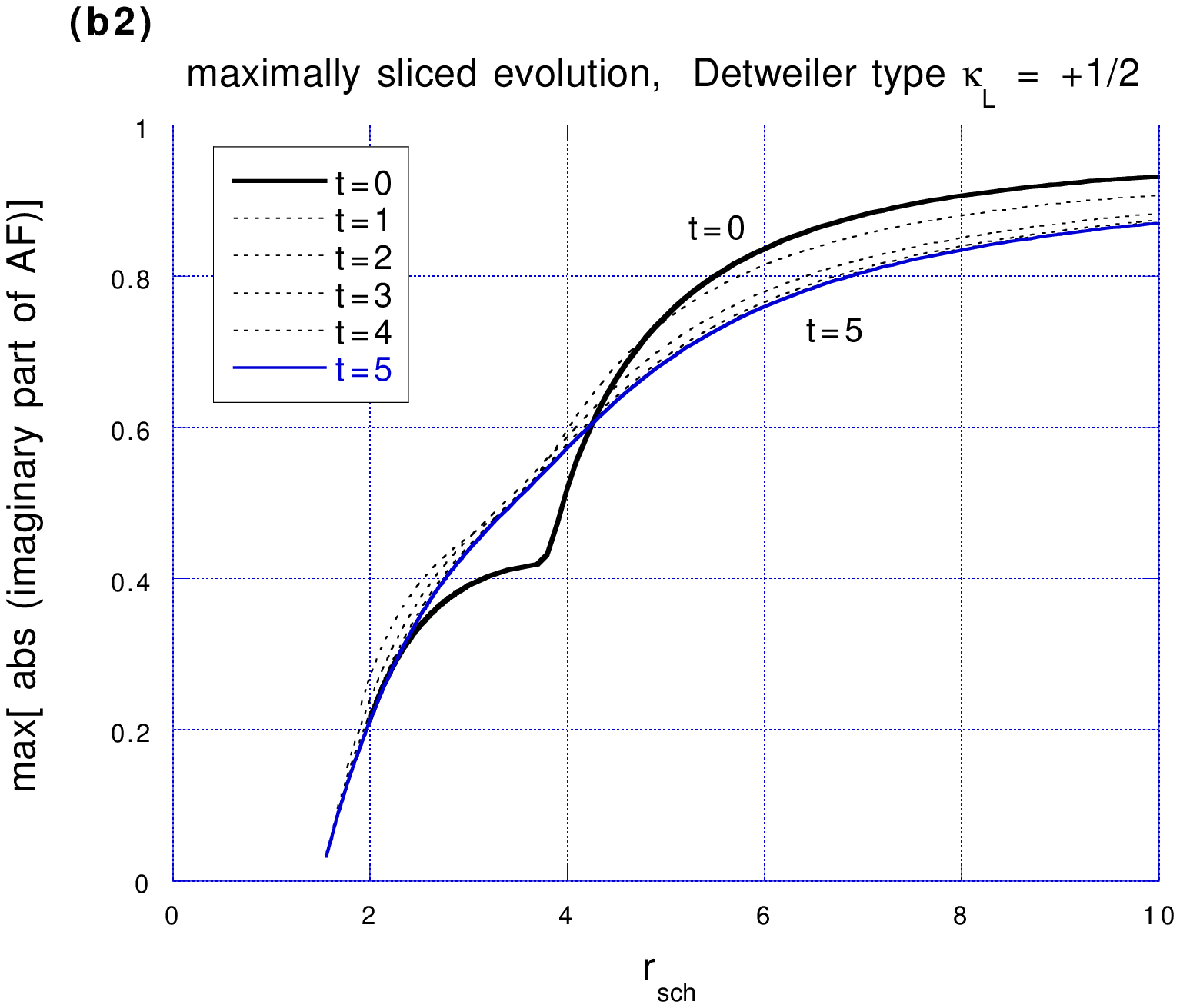} }
\end{picture}
\caption[quartic]{
Amplification factors of snapshots of maximally-sliced
evolving Schwarzschild spacetime. Fig (a1) and (a2) are of
the standard ADM formulation (real and imaginary parts, respectively), 
(b1) and (b2) are Detweiler's adjustment ($\kappa_L=+1/2$). 
Lines in (a1) and (b1) are the largest (positive) AF
on each time slice, while lines in (a2)  and (b2) are 
the maximum imaginary part of AF on each time slice.  
The time label $t$ in plots is $\bar{t}$ in Appendix \ref{appB}. 
The lines start at $r_{min}=2$ ($\bar{t}=0$) 
and $r_{min}=1.55$ ($\bar{t}=5$).
}
\label{fig7max}
\end{figure}

Fig.\ref{fig6penn} shows
the adjustment  No.4 in Table. \ref{table1}, which was 
used in the test of PennState group \cite{PennState_Sch1D}. 
The main difference from above is that the adjustment here is only 
for the radial component of the extrinsic curvature, $K_{rr}$.
The numerical experiments in \cite{PennState_Sch1D}
show better stability for positive 
$\kappa_\mu$, a fact that can be seen also from Fig.\ref{fig6penn}:
positive $\kappa_\mu$ produces negative real AFs. 
[See Fig.\ref{fig4adm}(b)(c) for the standard ADM case.]

Such kinds of test can be done with other combinations. 
In Table.\ref{table3}, we listed our results for more examples. 
The table includes the above results and is intended to extract the
contributions of each term in (\ref{adjADM1}) and (\ref{adjADM2}).
The effects of adjustments (of each $\kappa>0$ case) to AFs  are commented upon 
for each
coordinate system and for real/imaginary parts of AFs, respectively. 
These judgements are made at
the $r \sim O(10M)$ region on their $t=0$ slice.  
We hope this table will help 
further numerical improvements for the community. 

\subsection{in maximally-sliced evolving Schwarzschild spacetime}
\label{sec4d}

So far, our discussion is limited to one 3-hypersurface. 
Generally speaking, such an initial-value like analysis may not be 
enough to determine what combination of 
adjustments, coordinate system, and  gauge conditions are suitable
for the numerical evolution problem.  Here as the first further step, we
show our analysis of several snapshots of maximally sliced Schwarzschild
spacetime.  

The so-called maximal slicing condition, $K=0$, is one of the most widely used
gauge conditions to fix the lapse function, $\alpha$, during numerical evolution, 
since it has a feature of singularity avoidance. 
The condition will turn to an elliptic equation for $\alpha$. 
However, 
in the case of Schwarzschild spacetime, we can express a
maximally-sliced  hypersurface at an arbitrary time without full 
numerical time integration. 
The recipe is given by 
Estabrook et al \cite{estabrook}, and we introduce the procedure briefly
in Appendix \ref{appB}. 

By specifying a particular time $\bar{t}$, where $\bar{t}$
is supposed to express the time coordinate on 
a maximally sliced hypersurface (see detail in 
 Appendix \ref{appB}), we get metric components ($\alpha, \beta$ and 
$\gamma_{ij}$). 
 The procedure requires
us to solve ODE, but it is not a result of time integration.  
We then calculate  AFs as the previous ones. 

We show several snapshots in 
Fig.\ref{fig7max}.  
We picked up 3-slices of $\bar{t}=0,1,2,\cdots,5$
for  (a) the standard ADM and (b) Detweiler-type
adjustment for the Schwarzschild coordinate.  These initial data
($\bar{t}=0$) match those in Fig.\ref{fig1schadm}(a) and 
Fig.\ref{fig2schdet}(b), respectively.  
It is well known that the maximally-sliced hypersurfaces approach 
(or stop approaching to the singularity) $r_{min} \rightarrow (3/2)M$ as 
$\bar{t} \rightarrow \infty$. The snapshots here correspond to
$r_{min}=2.00, 1.90, 1.76, 1.66, 1.60$ and $1.56$ in unit $M$. 

The figures show that AFs are changing along their evolution although not
drastically, and also that their evolution 
behavior is dependent upon the choice of adjustments. 
The evolution makes the AFs' configuration converge, which is 
expected from the nature of maximal slicing. 

\section{Concluding Remarks}\label{Sec5}
Motivated by performing a long-term stable and accurate numerical simulation
of the Einstein equation,  
we proposed to adjust evolution equations by adding constraint terms
by analyzing the constraint propagation equations in advance. 
The idea is to construct an asymptotically constrained evolution system, 
which is robust against violation of the constraint. 
This method works even for the ADM formulation (which is not a hyperbolic system)
against flat background spacetime
\cite{ronbun3},  and here we applied the analyses to a 
curved spacetime, a spherically symmetric black hole spacetime. 

Recently, several numerical relativity groups report the effects of 
adjustments.  They are mostly searching for 
a suitable combination of multipliers 
through trial and error.  
We hope our discussion here helps
to understand the background mathematics systematically, though it may 
not be the perfect explanation. 
The main difference between our analysis and actual numerical studies is
that this is a local analysis only on the evolution equations. 
The actual numerical results are obtained under a certain treatment of 
the boundary conditions, 
the choice of numerical integration schemes and grid structures, 
and also depend on the accuracy of the initial data.
We think, however, 
that our proposal is an alternative to the hyperbolicity
classification in that it includes the non-principal part
of the evolution equations, and we expect that 
the discussion here will provide fundamental 
information on the stable formulation of the Einstein equations to 
the community.
Although we have not shown any numerical confirmations of our 
conjecture in this article, we remark that the amplification
factors explain the constraint violation behaviors quite well in the
Maxwell equations and in the Ashtekar version of the Einstein equation
on the flat background\cite{ronbun2}, and also that one of our examples
explains a successful case of Kelly et al \cite{PennState_Sch1D}. 

We presented a useful expression for analyzing ADM constraint propagation in 
general in Appendix \ref{appA}, and several analytic predictions
of the adjustments for the Schwarzschild spacetime in Table. \ref{table3}.  
We searched when and where the negative real or non-zero imaginary 
eigenvalues of the 
constraint propagation matrix appear, and how they depend on the choice
of coordinate system and adjustments.  
Our analysis includes the proposal of
Detweiler (1987), which is still the best one though it has a growing mode
of constraint violation near the horizon.  

We observed that the effects of adjustments depend on the choice of 
coordinate, gauge conditions, and also on its time evolution.  
Therefore our basic assumption of the constancy of the multipliers may be
better to  
be replaced with more general procedures in our future treatment.  
We have already started to study this issue by applying the recent development of 
the computational techniques in classical multi-body dynamical 
systems \cite{next}, and
hope that we can present some results soon elsewhere. 

\begin{acknowledgments}
We thank Y. Erigushi, T. Harada, M. Shibata, M. Tiglio for helpful comments. 
HS is supported by the special postdoctoral researchers program at
RIKEN. 
\end{acknowledgments}

\appendix
\section{General expressions of ADM constraint propagation equations}
\label{appA}
For the reader's convenience, we express here the
constraint propagation equations generally, considering the adjustments
to the evolution equations.
We repeat the necessary equations again in order for this appendix
to be read
independently.

\subsection{The standard ADM equations and  constraint propagations}
We start by analyzing the standard ADM system, that is,
with evolution equations
\begin{eqnarray}
\partial_t \gamma_{ij}&=&
-2\alpha K_{ij}+\nabla_i\beta_j+\nabla_j\beta_i, \label{Aadmevo1}
\\
\partial_t K_{ij} &=&
\alpha R^{(3)}_{ij}+\alpha K K_{ij}-2\alpha K_{ik}{K^k}_j
-\nabla_i\nabla_j \alpha
\nonumber \\&&
+(\nabla_i \beta^k) K_{kj} +(\nabla_j \beta^k) K_{ki}
+\beta^k \nabla_k K_{ij},  \label{Aadmevo2}
\end{eqnarray}
and constraint equations
\begin{eqnarray}
{\cal H}&:=& R^{(3)}+K^2-K_{ij}K^{ij}, \label{AadmCH} 
\\
{\cal M}_i&:=& \nabla_j {K^j}_i-\nabla_i K,  \label{AadmCM}
\end{eqnarray}
where $(\gamma_{ij}, K_{ij})$ are the induced three-metric
and the extrinsic
curvature, $(\alpha, \beta_i)$ are the lapse function
and the shift covector,
$\nabla_i$ is the covariant derivative adapted to $\gamma_{ij}$,
and $R^{(3)}_{ij}$ is the three-Ricci tensor.

The constraint propagation equations,
which are the time evolution equations
of the Hamiltonian constraint (\ref{AadmCH}) and
the momentum constraints (\ref{AadmCM}).

\paragraph{Expression using ${\cal H}$ and ${\cal M}_i$}
The constraint propagation equations
can be written as
\begin{eqnarray}
\partial_t {\cal H}&=&
\beta^j (\partial_j {\cal H})
+2\alpha K{\cal H}
-2\alpha \gamma^{ij}(\partial_i {\cal M}_j)
\nonumber \\&&
+\alpha(\partial_l \gamma_{mk})
   (2\gamma^{ml}\gamma^{kj}-\gamma^{mk}\gamma^{lj}){\cal M}_j
-4\gamma^{ij} (\partial_j\alpha){\cal M}_i,
\label{ACHproADM}
\\
\partial_t {\cal M}_i&=&
-(1/2)\alpha (\partial_i {\cal H})
-(\partial_i\alpha){\cal H}
+\beta^j (\partial_j {\cal M}_i)
\nonumber \\&&
+\alpha K {\cal M}_i
-\beta^k\gamma^{jl}(\partial_i\gamma_{lk}){\cal M}_j
+(\partial_i\beta_k)\gamma^{kj}{\cal M}_j.
\label{ACMproADM}
\end{eqnarray}
This is a suitable form to discuss hyperbolicity of the system.
The simplest derivation of (\ref{ACHproADM}) and (\ref{ACMproADM})
is by using the Bianchi identity, which
can be seen in Frittelli \cite{Fri-con}.

A shorter expression is available, e.g.
\begin{eqnarray}
\partial_t {\cal H}
&=&
  \beta^l \partial_l {\cal H} +2\alpha K{\cal H}
-2\alpha \gamma^{-1/2}\partial_l(\sqrt\gamma {\cal M}^l)
-4(\partial_l\alpha){\cal M}^l
\nonumber
\\
&=&
   \beta^l \nabla_l {\cal H} +2\alpha K{\cal H}
-2\alpha (\nabla_l {\cal M}^l)
-4(\nabla_l\alpha){\cal M}^l,
\label{YS35b}
\\
  \partial_t   {\cal M}_i &=&
-(1/2)\alpha(\partial_i {\cal H})
-(\partial_i \alpha) {\cal H}
+ \beta^l \nabla_l {\cal M}_i
+\alpha K {\cal M}_i
+ (\nabla_i \beta_l ) {\cal M}^l
\nonumber \\
&=&
-(1/2)\alpha(\nabla_i {\cal H})
-(\nabla_i \alpha) {\cal H}
+ \beta^l \nabla_l {\cal M}_i
+\alpha K {\cal M}_i
+ (\nabla_i \beta_l ) {\cal M}^l,
\label{YS36b}
\end{eqnarray}
or by using Lie derivatives along $\alpha n^\mu$, 
\begin{eqnarray}
\pounds_{\alpha n^\mu} {\cal H}
&=&
2\alpha K{\cal H}
-2\alpha \gamma^{-1/2}\partial_l(\sqrt\gamma {\cal M}^l)
-4(\partial_l\alpha){\cal M}^l,
\label{YS35alie}
\\
\pounds_{\alpha n^\mu}  {\cal M}_i &=&
-(1/2)\alpha(\partial_i {\cal H})
-(\partial_i \alpha) {\cal H}
+\alpha K {\cal M}_i.
\label{YS35blie}
\end{eqnarray}

\paragraph{Expression using $\gamma_{ij}$ and $K_{ij}$}
In order to check the effects of the adjustments in (\ref{Aadmevo1})
and (\ref{Aadmevo2}) to constraint propagation,
it is useful to re-express (\ref{ACHproADM}) and (\ref{ACMproADM})
using $\gamma_{ij}$ and $K_{ij}$.
By a straightforward calculation, we obtain an expression as
\begin{eqnarray}
\partial_t {\cal H} &=&
H_1^{mn}(\partial_t\gamma_{mn})
+H_2^{imn}\partial_i(\partial_t \gamma_{mn})
+H_3^{ijmn}\partial_i\partial_j (\partial_t \gamma_{mn})
+H_4^{mn}(\partial_t K_{mn}), \label{CHpro_new}
\\
\partial_t {\cal M}_i
&=&
M_1{}_i{}^{mn}(\partial_t \gamma_{mn})
+M_2{}_i{}^{jmn}\partial_j(\partial_t \gamma_{mn})
+M_3{}_i{}^{mn}(\partial_t K_{mn})
+M_4{}_i{}^{jmn} \partial_j(\partial_t K_{mn}), \label{CMpro_new}
\end{eqnarray}
where
\begin{eqnarray}
H_1^{mn}&:=&
-2R^{(3)}{}^{mn}
-\Gamma^p_{kj}\Gamma^k_{pi}\gamma^{mi}\gamma^{nj}
+\Gamma^m\Gamma^n
\nonumber \\ &&
+\gamma^{ij}\gamma^{np}(\partial_i\gamma^{mk})(\partial_j\gamma_{kp})
-\gamma^{mp}\gamma^{ni}(\partial_i\gamma^{kj})(\partial_j\gamma_{kp})
-2K K^{mn}+2K^n{}_jK^{mj},
\\
H_2^{imn}&:=&
-2\gamma^{mi}\Gamma^n
-(3/2)\gamma^{ij}(\partial_j\gamma^{mn})
+\gamma^{mj}(\partial_j\gamma^{in})
+\gamma^{mn}\Gamma^i,
\\
H_3^{ijmn}&:=&
-\gamma^{ij}\gamma^{mn}+\gamma^{in}\gamma^{mj},
\\
H_4^{mn}&:=&
2(K\gamma^{mn}-K^{mn}),
\\
M_1{}_i{}^{mn}&:=&
   \gamma^{nj}(\partial_i K^m{}_j)
-\gamma^{mj}(\partial_j K^n{}_{i})
+(1/2)(\partial_j\gamma^{mn})K^j{}_i
+\Gamma^n K^m{}_{i},
\\
M_2{}_i{}^{jmn}&:=&
-\gamma^{mj}K^n{}_i
+ (1/2) \gamma^{mn}K^j{}_i
+(1/2)  K^{mn}\delta^j_i,
\\
M_3{}_i{}^{mn}&:=&
-\delta^n_i \Gamma^m
-(1/2)(\partial_i\gamma^{mn}),
\\
M_4{}_i{}^{jmn}&:=&
\gamma^{mj}\delta^n_i -\gamma^{mn}\delta^j_i,
\end{eqnarray}
where we expressed $\Gamma^m=\Gamma^m_{ij}\gamma^{ij}$. 

\subsection{Adjustments}

Generally, we here write the adjustment terms to
(\ref{Aadmevo1}) and (\ref{Aadmevo2})
using (\ref{AadmCH}) and (\ref{AadmCM}) by the following combinations,
\begin{eqnarray}
&\mbox{adjustment term of }\quad
\partial_t \gamma_{ij}:& \quad
+P_{ij} {\cal H}
+Q^k{}_{ij}{\cal M}_k 
+p^k{}_{ij}(\nabla_k {\cal H})
+q^{kl}{}_{ij}(\nabla_k {\cal M}_l),
\label{AadjADM1}
\\
&\mbox{adjustment term of}\quad
\partial_t K_{ij}:& \quad
+R_{ij} {\cal H}
+S^k{}_{ij}{\cal M}_k 
+r^k{}_{ij} (\nabla_k{\cal H})
+s^{kl}{}_{ij}(\nabla_k {\cal M}_l),
\label{AadjADM2}
\end{eqnarray}
where $P, Q, R, S$ and $p, q, r, s$
are multipliers  (please do not
confuse $R_{ij}$ with
three Ricci curvature  that we write as $R^{(3)}_{ij}$).
We adjust them only using up to the first derivatives in order to make 
the discussion simple. 

By substituting the above adjustments into (\ref{CHpro_new}) and (\ref{CMpro_new}),
we can write the adjusted constraint propagation equations as
\begin{eqnarray}
\partial_t {\cal H} &=&
\mbox{(original terms)}
\nonumber \\&&
  + H_1^{mn}[P_{mn} {\cal H}+Q^k_{mn}{\cal M}_k
            +p^k{}_{mn}(\nabla_k {\cal H})
            +q^{kl}{}_{mn}(\nabla_k {\cal M}_l)]
\nonumber \\&&
+H_2^{imn}\partial_i [P_{mn} {\cal H}+Q^k_{mn}{\cal M}_k
            +p^k{}_{mn}(\nabla_k {\cal H})
            +q^{kl}{}_{mn}(\nabla_k {\cal M}_l)]
\nonumber \\&&
+H_3^{ijmn}\partial_i\partial_j [P_{mn} {\cal H}+Q^k_{mn}{\cal M}_k
            +p^k{}_{mn}(\nabla_k {\cal H})
            +q^{kl}{}_{mn}(\nabla_k {\cal M}_l)]
\nonumber \\&&
+H_4^{mn} [R_{mn} {\cal H}+S^k_{mn}{\cal M}_k
            +r^k{}_{mn} (\nabla_k{\cal H})
            +s^{kl}{}_{mn}(\nabla_k {\cal M}_l)], \label{CHpro_newhosei}
\\
\partial_t {\cal M}_i &=&
\mbox{(original terms)}
\nonumber \\&&
+ M_1{}_i{}^{mn} [P_{mn} {\cal H}+Q^k_{mn}{\cal M}_k
            +p^k{}_{mn}(\nabla_k {\cal H})
            +q^{kl}{}_{mn}(\nabla_k {\cal M}_l)]
\nonumber \\&&
+M_2{}_i{}^{jmn} \partial_j [P_{mn} {\cal H}+Q^k_{mn}{\cal M}_k
            +p^k{}_{mn}(\nabla_k {\cal H})
            +q^{kl}{}_{mn}(\nabla_k {\cal M}_l)]
\nonumber \\&&
+M_3{}_i{}^{mn}  [R_{mn} {\cal H}+S^k_{mn}{\cal M}_k
            +r^k{}_{mn} (\nabla_k{\cal H})
            +s^{kl}{}_{mn}(\nabla_k {\cal M}_l)]
\nonumber \\&&
+M_4{}_i{}^{jmn} \partial_j [R_{mn} {\cal H}+S^k_{mn}{\cal M}_k
            +r^k{}_{mn} (\nabla_k{\cal H})
            +s^{kl}{}_{mn}(\nabla_k {\cal M}_l)]. \label{CMpro_newhosei}
\end{eqnarray}
Here the ``original terms" can be understood either as 
(\ref{ACHproADM}) and (\ref{ACMproADM}), or as 
(\ref{CHpro_new}) and (\ref{CMpro_new}).
Therefore, for example, we can see that adjustments to 
$\partial_t \gamma_{ij}$ 
do not always keep 
the constraint propagation equations in the first order form,
due to their contribution in the third adjusted term in (\ref{CHpro_newhosei}).

We note that these expressions of constraint propagation equations are
equivalent when we include the cosmological constant and/or matter terms.


\section{Maximally slicing a Schwarzschild black hole} \label{appB}
The maximal slicing condition, $K=0$, is one of the most widely used
gauge conditions to fix the lapse function, $\alpha$, during numerical evolution, 
since it has a feature of singularity avoidance. 
The condition will turn into an elliptic equation for $\alpha$, such as 
\begin{equation}
\nabla^i\nabla_i\alpha = ( ~^{(3)\!}R + K^2 ) \alpha
=K_{ij}K^{ij} \, 
\alpha,
\end{equation}
where we used the Hamiltonian constraint in the second equality.  However, 
for the case of Schwarzschild spacetime, we can express a
maximally-sliced  hypersurface at an arbitrary time without full 
numerical integration.  
The recipe is given by 
Estabrook et al \cite{estabrook}, and we introduce the procedure here briefly. 

The goal is to express Schwarzschild geometry with a metric 
\begin{equation}
ds^2=(-\alpha^2 + A^{-1} \beta^2) d\bar{t}^2 + 2 \beta d\bar{t} dr + A dr^2
+ r^2 (d \theta^2 + \sin^2 \theta \, d\varphi^2),
\end{equation}
where $\alpha(\bar{t},r), \beta(\bar{t},r)$ and $A(\bar{t},r)$, and we suppose
$\bar{t}$=const. on the maximally-sliced hypersurface. 

The original Schwarzschild time coordinate, $t$, is now 
the function of $t(\bar{t}, r)$, with 
\begin{eqnarray}
\left. {\partial t \over \partial \bar{t} } \right|_{r=const.} 
&=& \alpha \sqrt{A}
\label{esta12}
\\
\left. {\partial t \over \partial r} \right|_{\bar{t}=const.} &=& 
- {\sqrt{A}  T \over r^2}{ 1\over 1 -  2M/r} 
\label{esta13}
\end{eqnarray}
where $T$ is an arbitrary function of $\bar{t}$, while $A$ is given 
\begin{equation}
A = (1-{2M \over r} + {T^2 \over r^4})^{-1}, \label{Aeq}
\end{equation}
using both Hamiltonian and momentum constraints. (\ref{esta13}) 
can be integrated as
\begin{equation}
t= - {T \over M} \int^{X(T)}_{M/r} {1 \over 1-2x}
{1\over \sqrt{1-2x+T^2 M^{-4}x^4}} dx,
\label{jikan}
\end{equation}
where $X(T)$ is required to be a smaller real root of the quartic equation,
\begin{equation}
1-2x+T^2 M^{-4}x^4 =0, \label{quartic}
\end{equation}
 in order to be consistent with (\ref{esta12}) under the boundary condition, 
\begin{equation}
\mbox{the~smoothness~across~the~Einstein-Rosen~bridge}, 
~ r=r_{min}, ~ \mbox{~at~} t=0, 
\end{equation}
that turns $\partial t / \partial r |_{r \rightarrow r_{min}} 
\rightarrow \infty$ and 
$T=T(r_{min})=\sqrt{r^3_{min} (2 M - r_{min})}$. 

By identifying $t \sim \bar{t}$ at spatial infinity, we get
\begin{equation}
\bar{t}= - {T \over M} \int^{X(T)}_{0} 
{1 \over 1-2x}{1\over \sqrt{1-2x+T^2 M^{-4}x^4}} dx,
\label{jikan2}
\end{equation}
where the integration across the pole at $x=1/2$ is taken in the 
sense of the principal value. 

In summary, if we specified a parameter $T$ ($0\le T < 3\sqrt{3}/4$ corresponds to
$0\le \bar{t} < \infty $), then we obtain the coordinate time $t$ and $\bar{t}$ 
{}from (\ref{jikan}) and (\ref{jikan2})
via (\ref{quartic}). Then we obtain the metric components by 
\begin{eqnarray}
\alpha (r,T) &=& \sqrt{1-{2M\over r}+ {T^2 \over r^4}}
\left[1+ {dT \over d\bar{t}} {1 \over M} \int_0^{M/r}
 {dx \over (1-2x+T^2 M^{-4}x^4)^{2/3}} \right], 
\label{B10}
\\
\beta (r,T) &=& \alpha A T r^{-2}, 
\end{eqnarray}
and (\ref{Aeq}), where ${dT / d\bar{t}} $ in (\ref{B10})
can be calculated using 
\begin{eqnarray}
\left(
{dT \over d\bar{t}} \right)^{-1}=
 { d\bar{t} \over dT } &=& \lim_{Y \rightarrow X}
\left[
{T^2 M^{-5}X^4 \over (1-2T^2 M^{-4}X^3)(2X-1)\sqrt{1-2Y+T^2M^{-4}Y^4}}
\right. \nonumber \\&& \quad
\left.
-{1\over M}\int^Y_0  {dx  \over (1-2x+T^2M^{-4}x^4)^{3/2}}
\right]
\end{eqnarray}
They are functions of $r$ and the minimum value of $r$ (that is, at a throat), 
$r_{min}$, is given by a larger real root of $r^4-2Mr^{3}+T^2=0$. 
Note that the fact $r_{min} \rightarrow (3/2) M$ when $\bar{t} \rightarrow \infty$ 
indicates the singularity avoidance property of maximal slicing condition. 
The comparison of this feature with the harmonic slicing condition 
is seen in \cite{GeyerHerold}. 


\newpage

\begin{table}[h]
\begin{center}
\begin{ruledtabular}
\begin{tabular}{ll||l|c|c|l}
No. & name & adjustments (non-zero part) & TRS  & 1st? & motivations 
\\ \hline \hline 
0 & standard ADM & 
no adjustments  & -- & yes &
\\ \hline 
1 & original ADM & 
$R_{ij}=\kappa_F \alpha \gamma_{ij}$ & yes & yes &
$\kappa_F=-1/4$ makes original ADM
\\ \hline 
2-a & Detweiler  & 
(\ref{Det1})-(\ref{Det4}),  $\kappa_L$  & no & no &
proposed by Detweiler \cite{detweiler}
\\ \hline 
2-P & Detweiler P-part  & 
$P_{ij} = -\kappa_L \alpha^3 \gamma_{ij}$    & no & no &
only use $P_{ij}$ term of Detweiler-type
\\ \hline 
2-S & Detweiler S-part   & 
$S^k{}_{ij} = \kappa_L \alpha^2 [3 (\partial_{(i}\alpha) \delta^k_{j)} 
- (\partial_l \alpha ) \gamma_{ij}\gamma^{kl}]$ & 
no & yes & 
only use $S^k{}_{ij}$ term of Detweiler-type
\\ \hline 
2-s & Detweiler s-part   & 
$s^{kl}{}_{ij}= \kappa_L \alpha^3 [
  \delta^k_{(i}\delta^l_{j)}-(1/3) \gamma_{ij}\gamma^{kl}]$ & no & no & 
only use $s^{kl}{}_{ij}$ term of Detweiler-type
\\ \hline 
3 & simplified Detweiler  & 
$P_{ij} = -\kappa_L \alpha \gamma_{ij}$ & no & no &
similar to above No.2-P, but $\alpha$ monotonic
\\ \hline 
4 & constant $R_{rr}$ & 
$R_{rr} = -\kappa_\mu \alpha $ & no & yes &
used by Penn State group \cite{PennState_Sch1D}
\\ 
\end{tabular}
\end{ruledtabular}
\end{center}
\caption{List of adjustments we tested and plotted. 
(See more cases in Table.\ref{table3}).   
The column of adjustments are nonzero multipliers
in terms of 
(\ref{adjADM1}) and (\ref{adjADM2}).
The column `TRS' indicates whether each adjusting term satisfies the 
time reversal symmetry or not
based on the standard Schwarzschild coordinate. 
(`No' is the candidate that makes asymmetric amplification 
factors (AFs).)
 The column `1st?' indicates whether each adjusting term breaks the first-order feature of 
the standard constraint propagation equation, (\ref{CHproADM}) and (\ref{CMproADM}). 
(`Yes' keeps the system first-order, `No' is the candidate of breaking hyperbolicity of 
constraint propagation.) 
}
\label{table1}
\end{table}

\begin{table}[h]
\begin{center}
\begin{ruledtabular}
\begin{tabular}{ll||l|l|l|l|l}
\multicolumn{2}{c||}{adjustment $\backslash$ coordinate } 
& 
\multicolumn{2}{c|}{standard (\ref{standardSch})} &
{isotropic (\ref{isoSch})} & iEF (\ref{iEFSch}) & PG (\ref{PGSch})
\\  \hline  
&        & exact & maximal & exact & $t_{iEF}=0$ &  $t_{PG}=0$
\\ \hline \hline 
0 & standard ADM & Fig.\ref{fig1schadm}(a) &  Fig.\ref{fig7max}(a) & 
Fig.\ref{fig4adm}(a)
&Fig.\ref{fig4adm}(b) & Fig.\ref{fig4adm}(c) 
\\ \hline 
1 & original ADM & Fig.\ref{fig1schadm}(b) &  &  & & 
\\ \hline 
2-a & Detweiler    & Fig.\ref{fig2schdet} &  Fig.\ref{fig7max}(b) &
Fig.\ref{fig5det}(a)
&Fig.\ref{fig5det}(b) & Fig.\ref{fig5det}(c) 
\\ \hline 
2-P & Detweiler P-part & Fig.\ref{fig3schdetpart}(a)  &  &&   &     
\\ \hline 
2-S & Detweiler S-part & Fig.\ref{fig3schdetpart}(b) &  &&   &     
\\ \hline 
2-s & Detweiler s-part & Fig.\ref{fig3schdetpart}(c) &  &&   &      
\\ \hline 
3 & simplified Detweiler & Fig.\ref{fig3schdetpart}(d) && 
&&
\\ \hline 
4 & constant $R_{rr}$ &  && 
&Fig.\ref{fig6penn}(a) & Fig.\ref{fig6penn}(b) 
\\  
\end{tabular}
\end{ruledtabular}
\end{center}
\caption{List of figures presented in this article. The adjustments are explained in 
Table. \ref{table1}. }
\label{table2}
\end{table}

\begin{table}[h]
\begin{center}
\begin{ruledtabular}
\begin{tabular}{l|c|ll|c||c|c|c|c|c}
No. &No. in & \multicolumn{2}{l|}{adjustment} &  1st? &
 \multicolumn{3}{c|}{Sch/iso coords.} & \multicolumn{2}{c}{iEF/PG coords.} 
\\
  & Table.\ref{table1}  &  &    &   & TRS  & real. & imag. &  real. & imag.  
\\ \hline \hline 
0 & 0 & -- & no adjustments  & yes & -- & -- & -- & -- & -- 
\\ \hline 
P-1 & 2-P &
$P_{ij}$ & $-\kappa_L \alpha^3 \gamma_{ij}$ & no & no
& makes 2 Neg. & not apparent & makes 2 Neg. & not apparent 
\\ \hline 
P-2 & 3 &
$P_{ij}$ & $-\kappa_L \alpha \gamma_{ij}$ & no & no  
& makes 2 Neg. & not apparent & makes 2 Neg. & not apparent 
\\ \hline 
P-3 & -  &
$P_{ij}$ & $P_{rr}=-\kappa$ or $P_{rr}=-\kappa \alpha$ & no &   no
& slightly enl.Neg.  & not apparent  & slightly enl.Neg.  & not apparent 
\\ \hline 
P-4 & -  &
$P_{ij}$ & $-\kappa \gamma_{ij}$ & no  &  no
& makes 2 Neg.  & not apparent  & makes 2 Neg. &  not apparent
\\ \hline 
P-5 & -  &
$P_{ij}$ & $-\kappa \gamma_{rr}$ & no &   no
& red. Pos./enl.Neg.  & not apparent  &red.Pos./enl.Neg.  & not apparent
\\ \hline 
Q-1  & - & 
$Q^k{}_{ij}$ 
& $\kappa \alpha \beta^k \gamma_{ij}$   & no &no  & N/A & N/A & 
$\kappa\sim1.35$ min. vals. & not
apparent
\\ \hline 
Q-2  & - & 
$Q^k{}_{ij}$ 
& $Q^r{}_{rr}=\kappa$   & no & yes & 
red. abs vals. & not apparent  & red. abs vals.  & not apparent
\\ \hline 
Q-3  & - & 
$Q^k{}_{ij}$ 
& $Q^r{}_{ij}=\kappa\gamma_{ij}$ or $Q^r{}_{ij}=\kappa \alpha \gamma_{ij}$  & no & yes &  
red. abs vals. & not apparent & enl.Neg. & enl. vals.
\\ \hline 
Q-4  & - & 
$Q^k{}_{ij}$ 
& $Q^r{}_{rr}=\kappa \gamma_{rr}$   & no & yes &  
red. abs vals. & not apparent   & red. abs vals. & not apparent
\\ \hline 
R-1 & 1 & $R_{ij}$ & $\kappa_F \alpha \gamma_{ij}$ & yes & yes 
& \multicolumn{2}{c|}{$\kappa_F=-1/4$ min. abs vals.}  &
\multicolumn{2}{c}{$\kappa_F=-1/4$
min. vals.}
\\ \hline 
R-2 & 4 & $R_{ij}$ & $R_{rr} = -\kappa_\mu \alpha $ or $R_{rr} = -\kappa_\mu$ & yes & no 
& not apparent &  not apparent & red.Pos./enl.Neg. & enl. vals.
\\ \hline 
R-3 & - & $R_{ij}$ & $R_{rr} = -\kappa \gamma_{rr} $ & yes & no 
& enl. vals.  & not apparent   & red.Pos./enl.Neg.  &  enl. vals.
\\ \hline 
S-1 & 2-S &
$S^k{}_{ij}$ & $\kappa_L \alpha^2 [3 (\partial_{(i}\alpha) \delta^k_{j)}
- (\partial_l \alpha ) \gamma_{ij}\gamma^{kl}]$ & yes & no  & not apparent & not apparent & 
not apparent & not apparent
\\ \hline 
S-2 & - & $S^k{}_{ij}$ & 
$\kappa \alpha \gamma^{lk} (\partial_l \gamma_{ij}) $ & yes & no
& makes 2 Neg.  & not apparent  & makes 2 Neg. & not apparent
\\ \hline 
p-1 & -  & 
$p^{k}{}_{ij}$ & $p^{r}{}_{ij}=-\kappa \alpha \gamma_{ij}$ & no  & no  
& red. Pos.   & red. vals.  & red. Pos.  & enl. vals.
\\ \hline 
p-2 & -  & 
$p^{k}{}_{ij}$ & $p^{r}{}_{rr}=\kappa \alpha$ & no  &   no
& red. Pos.  & red. vals.  & red.Pos/enl.Neg. & enl. vals.
\\ \hline 
p-3 & -  & 
$p^{k}{}_{ij}$ & $p^{r}{}_{rr}=\kappa \alpha \gamma_{rr}$ &  no &   no
& makes 2 Neg. & enl. vals. & red. Pos. vals. &  red. vals.
\\ \hline 
q-1 & -  & 
$q^{kl}{}_{ij}$ & $q^{rr}{}_{ij}=\kappa \alpha \gamma_{ij}$ & no   & no  
& $\kappa=1/2$ min. vals.  & red. vals.  & not apparent & enl. vals. 
\\ \hline 
q-2 & -  & 
$q^{kl}{}_{ij}$ & $q^{rr}{}_{rr}=-\kappa \alpha \gamma_{rr}$ & no  &  yes
& red. abs vals.  & not apparent  & not apparent & not apparent
\\ \hline 
r-1 & -  & 
$r^{k}{}_{ij}$ & $r^{r}{}_{ij}=\kappa \alpha \gamma_{ij}$ & no  &  yes
& not apparent & not apparent  & not apparent & enl. vals. 
\\ \hline 
r-2 & -  & 
$r^{k}{}_{ij}$ & $r^{r}{}_{rr}=-\kappa \alpha$ &  no  &  yes  
& red. abs vals. & enl. vals.  & red. abs vals. &  enl. vals.
\\ \hline 
r-3 & -  & 
$r^{k}{}_{ij}$ & $r^{r}{}_{rr}=-\kappa \alpha \gamma_{rr}$ & no  &  yes
& red. abs vals. & enl. vals. & red. abs vals. & enl. vals.
\\ \hline 
s-1 & 2-s  & 
$s^{kl}{}_{ij}$ & $\kappa_L \alpha^3 [
  \delta^k_{(i}\delta^l_{j)}-(1/3) \gamma_{ij}\gamma^{kl}]$ & no & no 
& makes 4 Neg. & not apparent & makes 4 Neg. & not apparent
\\ \hline 
s-2 & -  & 
$s^{kl}{}_{ij}$ & $s^{rr}{}_{ij}=-\kappa \alpha \gamma_{ij}$ & no  &  no
& makes 2 Neg.  & red. vals.  & makes 2 Neg. & red. vals.
\\ \hline 
s-3 & -  & 
$s^{kl}{}_{ij}$ & $s^{rr}{}_{rr}= - \kappa \alpha \gamma_{rr}$ & no  &  no
& makes 2 Neg.  & red. vals.  & makes 2 Neg. & red. vals.
\\ 
\end{tabular}
\end{ruledtabular}
\end{center}
\caption{List of adjustments we tested in the Schwarzschild spacetime.  
The column of adjustments are nonzero multipliers
in terms of (\ref{adjADM1}) and (\ref{adjADM2}).
The column  `1st?' and `TRS' are the same as Table. \ref{table1}. 
The effects to amplification factors (when $\kappa>0$) are commented for each coordinate system
and for real/imaginary parts of AFs, respectively. 
The `N/A' means that there is no effect due to the coordinate properties;  
`not apparent' means the adjustment does not change the AFs
effectively according to our conjecture;
`enl./red./min.' means enlarge/reduce/minimize,  and 
`Pos./Neg.' means positive/negative, respectively. 
These judgements are made
at the $r \sim O(10M)$ region on their $t=0$ slice.  
}
\label{table3}
\end{table}



\end{document}